\documentclass[preprint,showpacs,showkeys,preprintnumbers, amsmath,amssymb, aps, prfluids,floatfix,superscriptaddress]{revtex4-2}

\usepackage{color}
\usepackage{amsmath}
\usepackage{graphicx}% Include figure files
\usepackage{dcolumn}% Align table columns on decimal point
\usepackage{bm}% bold math
\usepackage{hyperref}
\usepackage{etoolbox}

\usepackage{soul}

\def\hh{\mathfrak{h}}
\def\abs#1{\left|#1\right|}

\begin{document}

\title{Bounded flows of dense gases}

\author{Sergiu Busuioc}
\email[]{sergiu.busuioc@e-uvt.ro}
\affiliation{Department of Physics, West University of Timişoara\\
Bd. Vasile P\^{a}rvan 4, 300223 Timişoara, Romania}
\affiliation{Institute for Advanced Environmental Research, West University of Timişoara\\
Bd. Vasile P\^{a}rvan 4, 300223 Timişoara, Romania}
\author{Victor Sofonea}
\email[]{sofonea@gmail.com}
\affiliation{Department of Physics, West University of Timişoara\\
Bd. Vasile P\^{a}rvan 4, 300223 Timişoara, Romania}
\affiliation{Center for Fundamental and Advanced Technical Research, Romanian Academy\\
Bd. Mihai Viteazul 24, 300223, Timişoara, Romania}
{\large }
\date{\today}

\begin{abstract}
Numerical solutions of the Enskog equation obtained employing a Finite-Difference Lattice Boltzmann (FDLB) and a Direct Simulation Monte Carlo (DSMC)-like particle method (PM) are systematically compared to determine the range of applicability of the simplified Enskog collision operator implemented in the Lattice Boltzmann framework. Three types of bounded flows of dense gases -- namely the Fourier, the Couette, and the Poiseuille flows -- are investigated for a wide range of input parameters. For low to moderate reduced density, the proposed FDLB model exhibits commendable accuracy for all bounded flows tested in this study, with substantially lower computational cost than the PM method.

\end{abstract}


\maketitle

\section{\label{sec:intro}Introduction}

In recent decades, the study of rarefied gas flows characterized by non-negligible values of the Knudsen number ($Kn$), which represents the ratio between the mean free path of fluid particles in a gas and the characteristic length of the flow domain, has yielded significant progress. These flows have been numerically studied using the Boltzmann equation, considering the fluid constituents as point particles. However, when the mean free path of the fluid particles becomes comparable to their size, the influence of the finite molecular size becomes crucial~\cite{FK72}. This scenario arises in various practical applications, such as gas extraction in unconventional reservoirs~\cite{WLRZ16,SPC17}, high-pressure shock tubes~\cite{PH01}, flows through microfabricated nanomembranes~\cite{HPWSAGNB06} and single-bubble sonoluminescence~\cite{BHL02}.

The Enskog equation offers a means to extend the kinetic theory description of fluids beyond the dilute-gas Boltzmann limit~\cite{enskog22, cowling70, FK72, K10, DBK2021}. Unlike the Boltzmann approach, the Enskog equation considers the finite size of gas molecules and incorporates the space correlations between colliding molecules, the molecular mutual shielding, as well as the reduction of the available volume. This equation can be solved numerically using probabilistic or deterministic methods, similar to the Boltzmann equation. Deterministic approaches, such as the Monte Carlo quadrature method~\cite{FS93}, the fast spectral method~\cite{WZR15,WLRZ16} and the Fokker-Planck approximation~\cite{SG17,SG19}, have been employed to solve the Enskog equation in recent years. Furthermore, probabilistic methods have emerged following the success of the Direct Simulation Monte Carlo method (DSMC)~\cite{B76} developed by Alexander et al.~\cite{AGA95}, Montanero et al.~\cite{MS96} and Frezzotti~\cite{F97b}.

Over time, the Enskog equation has been utilized to investigate the properties of dense gases composed of hard spheres near solid walls in micro- and nano-channels~\cite{D87,DM97,F97b,F99,NFSJMH06,SGLBZ20,WGLBZ23}. Its extension to weakly attracting hard-sphere systems has proven successful in describing liquid-vapor flows of monoatomic~\cite{FGL05,KKW14,FBG19,BGLS20} and polyatomic fluids~\cite{Bruno2019,BG20} or mixtures~\cite{KSKFW17}, the formation and the rupture of liquid menisci in nanochannels~\cite{BFG15}, as well as the growth/ collapse of spherical nano-droplets/bubbles~\cite{BFG23}.

Although the aforementioned methods are known for their reliability and accuracy, their high computational costs often make them impractical for various applications. To address this limitation, a convenient approach is to simplify the non-local Enskog collision integral by expanding it into a Taylor series around a specific point $\bm{x}$ in the coordinate space. The first term of this expansion corresponds to the conventional Boltzmann collision operator. Additionally, the second term can be further simplified by replacing the distribution function with the local equilibrium distribution function, a procedure which holds when the fluid is close to equilibrium~\cite{cowling70,K10}. This simplification has been employed in Lattice Boltzmann (LB) models to investigate non-ideal gases~\cite{L98,L00,MM08} and multiphase flows by incorporating long-range attractive forces~\cite{HD02}.

More recently, the simplified Enskog collision operator has been successfully implemented in various solvers, including the discrete velocity method~\cite{WWHLLZ20}, the discrete unified gas kinetic scheme (DUGKS)~\cite{CWWC23}, the double-distribution LB model~\cite{HWA21} and the discrete Boltzmann method~\cite{ZXQWW20, GXLLSS22, B23}. By employing the simplified Enskog collision operator, these solvers offer computationally efficient alternatives for investigating micro-scale flow phenomena while maintaining reasonable accuracy.

In this paper, the recently introduced Lattice Boltzmann model for dense gas flows~\cite{B23}, based on the full-range Gauss-Hermite quadrature, is further developed to account for the bounded flow of the Enskog gas between parallel walls. In order to tackle the wall-induced discontinuity, which becomes effective at non-negligible values of the Knudsen number, the half-range Gauss-Hermite quadrature method~\cite{AS16a, AS16b, AS18, AS2019} is used to reduce the numerical errors, as well as the computational costs.

Throughout the paper, we use the non-dimensionalization procedure based on the reference quantities described in Ref.~\cite{AS18}: $L_{\text{ref}}$ (length), $n_{\text{ref}}$ ( particle number density) and $T_{\text{ref}}$ (temperature). Accordingly, the reference momentum is defined as $p_{\text{ref}}=\sqrt{m_{\text{ref}} k_B T_{\text{ref}}}$ and the reference time is given by $t_{\text{ref}} = m_{\text{ref}} L_{\text{ref}}/p_{\text{ref}}$, where $m_{\text{ref}}$ represents the mass of a fluid particle.

This paper is organized as follows. In sec. \ref{sec:enskog}, the simplified Enskog equation is presented along with the Shakhov collision term. The Finite Difference Lattice Boltzmann (FDLB) model used to numerically solve the simplified Enskog equation when the flow domain is bounded by parallel walls is introduced in Sec.~\ref{sec:fdlb}. This model relies on half-range Gauss-Hermite quadratures in order to account for the boundary-induced discontinuities. The computer simulation results are reported in Sec.~\ref{sec:results}, which has three subsections dedicated to the Fourier, the Couette, and the Poiseuille flow, respectively. We conclude the paper in Sec.\ref{sec:conclusions}.

For the convenience of the reader, further information about the full-range Gauss-Hermite quadrature and a comparison to the half-range quadrature results is presented in Appendix~\ref{appendix:fullrange}, the heat flux components evaluation in the PM and the local maxima in the total heat flux are discussed in Appendix~\ref{appendix:heat_flux}, while Appendix~\ref{appendix:pm} briefly presents the particle method (PM) of solving the Enskog equation~\cite{F97b}, which is systematically used to validate the FDLB results for the bounded flows listed above.

\section{\label{sec:enskog}The Enskog equation}

The Enskog equation, proposed in 1922~\cite{enskog22}, describes the evolution of a system consisting of rigid spherical molecules. Unlike Boltzmann's equation, which assumes molecules as point-like particles subjected to local collisions, Enskog's equation considers the volume of fluid particles (i.e., molecules). This volume restricts the free movement space available to each particle, leading to an increased number of collisions. Additionally, the collisions between particles are non-local, occurring when the centers of the two colliding molecules are separated by one molecular diameter. The Enskog equation can be written as follows~\cite{cowling70, K10, DBK2021}:
\begin{equation}\label{eq:enskog}
 \frac{\partial f}{\partial t}+\frac{\bm{p}}{m} \cdot\bm{\nabla}_{\bm{x}} f + \bm{F}\cdot\bm{\nabla}_{\bm{p}} f =J_E
\end{equation}
where $m$ denotes the particle mass, $\bm{F}=m\bm{a}$ represents the external body force and $f(\bm{x},\bm{p},t)$ is the single-particle distribution function. At time $t$, the distribution function $f$ provides the number of particles located within the phase space volume $d\bm{x}d\bm{p}$ centered at the point $(\bm{x},\bm{p})$. The right-hand side of the equation is given by the Enskog collision operator $J_E$, expressed as:
\begin{multline}\label{eq:collision_integral}
 J_E=\int \left\{ \chi\left({\bm{x}}+\frac{\sigma}{2}{\bm{k}}\right)f({\bm{x}},\bm{p^*})f({\bm{x}} + \sigma {\bm{k}},\bm{p_1^*}) \right.\\ -
    \left.   \chi\left({\bm{x}}-\frac{\sigma}{2}{\bm{k}}\right)f({\bm{x}},\bm{p})f({\bm{x}} - \sigma {\bm{k}},\bm{p_1}) \right\}\sigma^2
    ({\bm{p_r}}\cdot{\bm{k}}) d{\bm{k}}d{\bm{p_1}}
\end{multline}
In the above equation, $\sigma$ represents the molecular diameter, $\bm{p_r}=\bm{p_1}-\bm{p}$ is the relative momentum, and $\bm{k}$ is the unit vector specifying the relative position of the two colliding particles. The time dependence of the distribution function is omitted for brevity.

The contact value of the pair correlation function $\chi$ incorporates the effect of the molecular diameter $\sigma$ on the collision frequency. In the standard Enskog theory (SET), $\chi \equiv \chi_{\text{\tiny SET}}$ is evaluated at the contact point of two colliding particles in a fluid assumed to be in uniform equilibrium \cite{K10}. An approximate but accurate expression for $\chi_{\text{\tiny SET}}$, namely:
\begin{equation}\label{eq:chi}
 \chi_{\text{\tiny SET}}[n]=\frac{1}{nb}\left(\frac{P^{hs}}{n k_B T}-1\right)=\frac{1}{2}\frac{2-\eta}{(1-\eta)^3},
\end{equation}
is derived from the equation of state (EOS) for the hard-sphere fluid
\begin{equation}\label{eq:eos}
 P^{hs}=nk_B T\frac{1+\eta+\eta^2-\eta^3}{(1-\eta)^3}
\end{equation}
proposed by Carnahan and Starling~\cite{CS69}. Here $n$ represents the particle number density, $\eta=b \rho /4$ is the reduced particle density ($b=2\pi\sigma^3/3m$), $P^{hs}$ is the pressure of the hard-sphere fluid, $k_B$ is the Boltzmann constant and $T$ is the temperature. The square brackets in Eq.~\eqref{eq:chi} indicate a functional dependence.

In the revised Enskog theory (RET), the fluid is considered to be in a non-uniform equilibrium state~\cite{K10, DBK2021, vBE73}, hence the particle number density is position dependent. In this case, an effective approximation for the radial distribution function is obtained using the Fischer-Methfessel (FM) prescription~\cite{FM80}. This approach involves the replacement in Eq.\eqref{eq:chi} of the actual value of the particle number density $n$ with the average particle density $\overline{n}$ computed over a spherical volume of radius $\sigma$, centered at $\bm{x}-\frac{\sigma}{2} {\bm{k}}$:
\begin{equation}\label{eq:chi_ret}
\chi_{\mbox{\tiny RET-FM}}\left[n\Big(\bm{x}-\frac{\sigma}{2} {\bm{k}}\Big)\right]=\chi_{\mbox{\tiny SET}}\left[\overline{n}\Big(\bm{x}-\frac{\sigma}{2} {\bm{k}}\Big)\right].
\end{equation}
The average particle density $\overline{n}$ is given by:
\begin{equation}
\overline{n}(\bm{x})=\frac{3}{4\pi \sigma^3}\int_{\mathbb{R}^3} n(\bm{x}')w(\bm{x},\bm{x}')\,d\bm{x}',\quad
w(\bm{x},\bm{x}')=\left\{
\begin{array}{cc}
1, &\qquad \|\bm{x}'-\bm{x}\|<\sigma, \\
0, & \qquad \|\bm{x}'-\bm{x}\|\geq\sigma.
\end{array}
\right.
\end{equation}
In the rest of the paper the subscript RET-FM and the functional dependence of $\chi_{\mbox{\tiny RET-FM}}[n(\bm{x}-\frac{\sigma}{2} {\bm{k}})]$ will be omitted for brevity.

The Enskog collision operator $J_E$ in Eq.~\eqref{eq:collision_integral} can be seen as a generalization of the Boltzmann collision operator to account for particles with spatial extent. When the molecular diameter $\sigma$ approaches zero, the contact value of the pair correlation function approaches unity ($\chi\rightarrow 1$), which recovers the Boltzmann collision operator~\cite{cowling70,K10}.

\subsection{\label{sec:simpl_enskog}Simplified Enskog collision operator}

Assuming that the contact value of the pair correlation function $\chi$ and the distribution functions $\{f^*\equiv f({\bm{x}},\bm{p^*}),f_1^*\equiv f({\bm{x}} + \sigma {\bm{k}},\bm{p_1^*}),f\equiv f({\bm{x}} ,\bm{p}) ,f_1\equiv f({\bm{x}} - \sigma {\bm{k}},\bm{p_1})\}$, which appear in the Enskog collision integral $J_E$ given in Eq.(\ref{eq:collision_integral}), are smooth around the contact point $\bm{x}$, we can approximate these functions using a Taylor series expansion around $\bm{x}$. The simplified Enskog collision operator is obtained after retaining the resulting expansion of $J_E\approx\, J_0+J_1$ up to first-order gradients, namely~\cite{cowling70,K10}:
\begin{eqnarray}
 J_0 \equiv J_0[f] &=& \chi\int (f^*f_1^*-ff_1)\sigma^2({\bm{p_r}}\cdot{\bm{k}}) d{\bm{k}}d{\bm{p_1}}\\
 J_1 \equiv J_1[f] &=& \chi\sigma\int\bm{k}(f^*\bm{\nabla}f_1^*+f\bm{\nabla} f_1)\sigma^2({\bm{p_r}}\cdot{\bm{k}}) d{\bm{k}}d{\bm{p_1}}\nonumber\\
 &+&\frac{\sigma}{2}\int\bm{k}\bm{\nabla}\chi(f^*f_1^* + ff_1)\sigma^2({\bm{p_r}}\cdot{\bm{k}}) d{\bm{k}}d{\bm{p_1}}
\end{eqnarray}
The functions $f^*,f_1^*,f,f_1$ and $\chi$ in the two equations above are evaluated at the point ${\bm{x}}$.

The term $J_0[f]$ corresponds to the conventional collision term of the Boltzmann equation multiplied by $\chi$ and is treated as such by applying the relaxation time approximation. In this study, we employ the Shakhov collision term~\cite{shakhov68a,shakhov68b}:
\begin{equation}\label{eq:j0}
J_0[f]=-\frac{1}{\tau}(f-f^S),
\end{equation}
where $\tau$ represents the relaxation time and $f^S$ is the equilibrium Maxwell-Boltzmann distribution multiplied by a correction factor~\cite{shakhov68a,shakhov68b,GP09,ASS20}:
\begin{equation}
f^S=f^{\text{\tiny MB}}\left[1 + \frac{1-\text{Pr}}{P_i k_B T}\left( \frac{\bm{\xi}^2}{5 m k_B T}-1 \right)\bm{\xi}\cdot \bm{q} \right]
\end{equation}
The Maxwell-Boltzmann distribution $f^{\text{\tiny MB}}$ is defined as
\begin{equation}
f^{\text{\tiny MB}}=\frac{n}{(2 m \pi k_B T)^{3/2}}\exp{\left(-\frac{\bm{\xi}^2}{2 m k_B T}\right)}
\end{equation}
and the heat flux $\bm{q}$ is obtained using:
\begin{equation}
\bm{q}=\int d^3p f \frac{\bm{\xi}^2}{2m}\frac{\bm{\xi}}{m},
\end{equation}
where $\bm{\xi}=\bm{p}-m\bm{u}$ represents the peculiar momentum, $\text{Pr}=c_P\mu/\lambda$ denotes the Prandtl number, $c_P=5k_B/2m$ is the specific heat at constant pressure, $\mu$ is the shear viscosity, $\lambda$ is the thermal conductivity and $P_i=\rho R T=n k_B T$ is the ideal gas equation of state, with $R$ being the specific gas constant.
It is important to note that although the Shakhov model does not guarantee non-negativity of the correction factor and the H-theorem has not been proven, the model has been successfully implemented and its accuracy has been tested through comparisons with experimental~\cite{S02,S03,GP09} or DSMC~\cite{AS18, ZXZCW19, ASS20, TWS20} results.

The second term of $J_E$, denoted as $J_1[f]$, can be approximated by replacing the distribution functions ($f^*,f_1^*,f,f_1$) with their corresponding equilibrium distribution functions. By using $f^{*,{\text{\tiny MB}}}f_{1}^{*,\text{\tiny MB}}=f^{\text{\tiny MB}}f^{\text{\tiny MB}}_1$ and integrating over $\bm{k}$ and $\bm{p_1}$, we obtain~\cite{cowling70,K10}:
\begin{multline}\label{eq:j1}
 J_1[f]\approx J_1[f^{\text{\tiny MB}}]=-b \rho \chi f^{\text{\tiny MB}} \left\{\bm{\xi}\cdot\left[\bm{\nabla}\ln(\rho^2 \chi T)+\frac{3}{5}\left(\zeta^2-\frac{5}{2}\right)
 \bm{\nabla}\ln T\right]\right.\\
 \left. + \frac{2}{5}\left[ 2\bm{\zeta}\bm{\zeta}\bm{:\nabla u} + \left(\zeta^2-\frac{5}{2} \right)\bm{\nabla\cdot u} \right]
 \right\}
\end{multline}
where $\bm{\zeta}=\bm{\xi}/\sqrt{2RT}$. By incorporating these approximations, the Enskog equation Eq.~\eqref{eq:enskog} can be expressed as:
\begin{equation}\label{eq:enskog_approx}
\frac{\partial f}{\partial t}+\frac{\bm{p}}{m}\bm{\nabla}_{\bm{x}}f + \bm{F}\cdot\bm{\nabla}_{\bm{p}} f= -\frac{1}{\tau}(f-f^S)+J_1[f^{\text{\tiny MB}}]
\end{equation}

The macroscopic quantities can be determined by evaluating the corresponding moments of the distribution function:

\begingroup
\renewcommand*{\arraystretch}{1.25}
\begin{equation}
 \begin{pmatrix}
 n \\ \rho\bm{u} \\ \frac{3}{2} n k_B T \\ {\bm{\Pi}}^{kin}\\ {\bm{q}}^{kin}
 \end{pmatrix} =\int d^3 p
 \begin{pmatrix}
 1 \\ \bm{p} \\ \frac{\bm{\xi}^2}{2m}\\ \frac{\bm{\xi}\bm{\xi}}{m}\\ \frac{\bm{\xi}\bm{\xi}^2}{2m^2}
 \end{pmatrix} f
\end{equation}
\endgroup
where $\rho=mn$, while ${\bm{\Pi}}^{kin}$ and ${\bm{q}}^{kin}$ denote the kinetic stress tensor and the kinetic heat flux, respectively. Multiplying the Enskog equation Eq.\eqref{eq:enskog} by the collision invariants $1,{\bm{p}}$, and ${\bm p}^2 /2m$, and integrating over the momentum space, we obtain the following conservation equations for mass, momentum, and energy~\cite{K10}:
\begin{subequations}
\begin{align}
 \frac{D\rho}{Dt}+\rho\nabla \cdot\bm{u}&=0\\
 \rho \frac{D {\bm{u}}}{Dt}+\nabla P&=-\nabla\cdot{\bm{\Pi}} \\
 \rho\frac{De}{Dt}+P\nabla\cdot\bm{u}&=-\nabla\cdot \bm{q}+{\bm{\Pi}}\bm{:}\nabla\bm{u}
 \end{align}
\end{subequations}
Here, $D/Dt=\partial_t+\bm{u}\cdot\nabla$ represents the material derivative, and $P=P_i(1+b\rho\chi)$ denotes the equation of state for a non-ideal gas. The heat flux $\bm{q}$ and the viscous part of the stress tensor $\bm{\Pi}$ are given by:
\begin{subequations}\label{eq:heatstress}
\begin{eqnarray}
&\bm{q}=-\lambda\nabla T, \\
&\hspace{-10pt}\bm{\Pi}=-\mu_v\mathcal{I}\bm{\nabla} \cdot \bm{u}-\mu\left(\bm{\nabla u} + (\bm{\nabla u})^T-\frac{2}{3}\mathcal{I}\bm{\nabla} \cdot \bm{u}\right)
\end{eqnarray}
\end{subequations}
where $\mathcal{I}$ represents the identity matrix. The bulk viscosity $\mu_v$, the shear viscosity $\mu$ and the thermal conductivity $\lambda$, which appear in Eqs.~({\ref{eq:heatstress}), are given by~\cite{K10}:
\begin{eqnarray}
\mu_v & = & \frac{16}{5\pi}\mu_0 b^2 \rho^2 \chi\, , \label{eq:bviscosity} \\
 \mu = \tau P_i & = & \mu_0 \left[\frac{1}{\chi}+\frac{4}{5}(b\rho)+\frac{4}{25}\left(1+\frac{12}{\pi}\right)(b\rho)^2\chi\right]\, , \label{eq:viscosity} \\
 \lambda = \frac{5k_B}{2m}\frac{\tau P_i}{\text{Pr}} & = & \lambda_0 \left[\frac{1}{\chi}+\frac{6}{5}(b\rho)+\frac{9}{25}\left(1+\frac{32}{9\pi}\right)(b\rho)^2\chi\right]\, .
\end{eqnarray}
In these equations, $\mu_0$ and $\lambda_0$ represent the viscosity coefficient and thermal conductivity for hard-sphere molecules at temperature $T$, namely~\cite{K10}:
\begin{equation}
\mu_0=\frac{5}{16\sigma^2}\sqrt{\frac{m k_B T}{\pi}},\quad \lambda_0=\frac{75 k_B}{64 m\sigma^2}\sqrt{\frac{m k_B T}{\pi}}
\end{equation}
For a dense gas, the Prandtl number $\text{Pr}$ is expressed as~\cite{K10}:
\begin{equation}\label{eq:prandtl}
\text{Pr}=\frac{2}{3} \,\frac{1+\frac{4}{5}b\rho\chi+\frac{4}{25}\left(1+\frac{12}{\pi}\right)(b\rho\chi)^2}{1+\frac{6}{5}b\rho\chi+\frac{9}{25}\left(1+\frac{32}{9\pi}\right)(b\rho\chi)^2}
\end{equation}
The dilute limit corresponds to $\text{Pr}=2/3$.
The Chapman-Enskog expansion of Eq.~\eqref{eq:enskog_approx} provides relationships between the relaxation time $\tau$ and the transport coefficients. In this context, the relaxation time $\tau$ is expressed as:

\begin{equation}\label{eq:tau}
\tau=\frac{\mu}{P_i}
\end{equation}

The quantity $\mu$ encompasses both kinetic and potential contributions, which account for the flow of molecules and the collisional effects on the transfer of momentum and energy in the gas~\cite{cowling70,K10}. The relaxation time approximation effectively captures the collisional transfer resulting from non-local molecular collisions. It is worth noting that the viscosity of a dense gas with a fixed reduced density $\eta$ can be adjusted by varying the molecular diameter $\sigma$ and the number density $n$.

The Knudsen number {is defined as the ratio of the mean free path and a characteristic length (in our case the channel width):
\begin{equation}
 Kn=\frac{\lambda}{L}=\frac{1}{\sqrt{2}\pi\sigma^2 n_0 \chi(n_0) L}=\frac{1}{6\sqrt{2} \eta_0 \chi(\eta_0) R}
\end{equation}
where $R=L/\sigma$ is the confinement ratio~\cite{WZR15,SGLBZ20,WGLBZ23}.

This paper primarily concentrates on benchmarking the Fourier, Couette, and Poiseuille flow cases. In these scenarios, the steady flow either lacks bulk motion or exhibits motion perpendicular to the direction in which gas density varies. Consequently, the bulk viscosity does not have an impact in these cases.

The model equations involving the simplified Enskog collision operator $J_1$ Eq.~\eqref{eq:j1} are formulated using only a limited number of low-order derivatives, resulting in the omission of information contained in higher-order terms. As a consequence, the high-order information excluded in $J_1$, which is not present in the collisional momentum and energy transfer, is reintroduced in the kinetic transfer of momentum and energy through the relaxation time~\eqref{eq:tau} and the Prandtl number~\eqref{eq:prandtl} in the collision term $J_0[f]$~\eqref{eq:j0}. This ensures that the total stress tensor and heat flux derived from the current kinetic model align with those obtained from the Enskog equation, at least up to the first-order approximation\cite{K10,WGLBZ23}. As such, when we will compare the heat fluxes in the Fourier, Couette, and Poiseuille setups, the simulation results obtained using the PM will contain the total heat flux $q_x$, the kinetic as well as the potential contributions defined in Appendix~\ref{appendix:heat_flux}.

\subsection{\label{sec:reduced_dist}Reduced distributions}

In the channel flows examined in this paper, the dynamics along the $z$ direction is straightforward. Furthermore, in the heat transfer (i.e., Fourier) problem, also the dynamics along the y-axis is straightforward. In this context, it is advantageous to integrate out the trivial degrees of freedom in the momentum space at the level of the model equation.

\subsubsection{\label{sec:reduced_dist_1D}1D flows}

In the Fourier flow, the dynamics along the $y$ and $z$ directions is trivial. After integrating along these Cartesian axes, two reduced distribution functions, namely $\phi$ and $\theta$, are introduced~\cite{LZ04,GP09,MWRZ13,AS18,AS2019,BA19}:
\begin{align}
 \phi_{\text{1D}}({\bm x},p_x,t)&=\int dp_y dp_z f({\bm x},{\bm p},t),\\
 \theta_{\text{1D}}({\bm x},p_x,t)&= \int dp_y dp_z \frac{p_y^2+p_z^2}{m}f({\bm x},{\bm p},t)
\end{align}
In this case, the macroscopic quantities are given by:
\begin{subequations}\label{eq:macro1D}
\begingroup
\renewcommand*{\arraystretch}{1.2}
\begin{align}
 \begin{pmatrix}
 n \\ \rho u_x \\ \Pi_{xx}
 \end{pmatrix} &=\int d p_x
 \begin{pmatrix}
 1 \\ p_x \\ \frac{\xi_x^2}{m}
 \end{pmatrix} \phi_{\text{1D}},\\
 \begin{pmatrix}
 \frac{3}{2}n k_B T \\ q_x
 \end{pmatrix} &=\int d p_x
 \begin{pmatrix}
 1 \\ \frac{\xi_x}{m}
 \end{pmatrix} \left(\frac{\xi_x^2}{2m}\phi_{\text{1D}}+\frac{1}{2}\theta_{\text{1D}}\right)
\end{align}
\endgroup
\end{subequations}
and the evolution equations for the reduced distribution functions become:
\begingroup
\renewcommand*{\arraystretch}{1.25}
\begin{equation}\label{eq:evolution_1D}
 \frac{\partial}{\partial t}
 \begin{pmatrix}
 \phi_{\text{1D}}\\ \theta_{\text{1D}}
 \end{pmatrix}
+ \frac{p_x}{m}\frac{\partial}{\partial x}
\begin{pmatrix}
 \phi_{\text{1D}}\\ \theta_{\text{1D}}
 \end{pmatrix}
= -\frac{1}{\tau} \begin{pmatrix}
 \phi_{\text{1D}}-\phi_{\text{1D}}^S\\ \theta_{\text{1D}}-\theta_{\text{1D}}^S
 \end{pmatrix}
 + \begin{pmatrix}
 J_1^{\phi_{\text{1D}}}\\ J_1^{\theta_{\text{1D}}}
 \end{pmatrix}
 \end{equation}
\endgroup
In the above equations, $\phi_{\text{1D}}^S$ and $\theta_{\text{1D}}^S$ are given by:
\begin{align}
 \phi^S=f_x^{\text{\tiny MB}}\left[1 + \frac{1-\text{Pr}}{5P_i m k_B T}\left( \frac{\xi_x^2}{m k_B T}-3 \right)\xi_x q_x \right],\\
 \theta^S = 2 k_B T f_x^{\text{\tiny MB}}\left[1 + \frac{1-\text{Pr}}{5P_i m k_B T}\left( \frac{\xi_x^2}{m k_BT}-1 \right)\xi_x q_x \right]
 \end{align}
where
 \begin{equation}\label{eq:f_mb}
 f_x^{\text{\tiny MB}}=\frac{n}{(2 m \pi k_B T)^{1/2}}\exp{\left(-\frac{\xi_x^2}{2 m k_B T}\right)}
 \end{equation}
while the first order corrections $J_1^\phi$ and $J_1^\theta$ are:
 \begin{subequations}\label{eq:j1_expanded_1D}
 \begin{multline}
 J_1^{\phi_{\text{1D}}}=-\left[ \xi_x\partial_x\ln\chi + 2\xi_x\partial_x\ln\rho + %%\frac{3}
{5}\left(\frac{\xi_x^2}{m k_B T}-1 \right)\partial_x u_x \right. \\ \left.+
\frac{3}{10}\left(\frac{\xi_x^3}{m^2 k_B T}+\frac{\xi_x}{3m} \right)\partial_x\ln T
 \right]f_{x}^{\text{\tiny MB}}b\rho\chi
 \end{multline}
\begin{multline}
 J_1^{\theta_{\text{1D}}}=-\left[ \xi_x\partial_x\ln\chi + 2\xi_x\partial_x\ln\rho +
 \frac{3}{5}\left(\frac{\xi_x^2}{m K_B T}-\frac{1}{3} \right)\partial_x u_x \right. \\ \left.+
 \frac{3}{10}\left(\frac{\xi_x^3}{m^2 k_B T}+\frac{7\xi_x}{3m} \right)\partial_x\ln T
 \right]2 m k_B T f_{x}^{\text{\tiny MB}}b\rho\chi
\end{multline}
\end{subequations}

\subsubsection{\label{sec:reduced_dist_2D}2D flows}

To reduce the computational costs when simulating the Couette and the Poiseuille flows, where only the dynamics along the $z$ direction is trivial, we integrate along this direction in the momentum space and get the reduced distribution functions~\cite{LZ04,GP09,MWRZ13,AS18,AS2019,BA19}:
\begin{align}
 \phi_{\text{2D}}({\bm x},p_x,p_y,t)&=\int dp_z f({\bm x},{\bm p},t),\\
 \theta_{\text{2D}}({\bm x},p_x,p_y,t)&= \int dp_z \frac{p_z^2}{m}f({\bm x},{\bm p},t)
\end{align}
This way, the macroscopic quantities are given by:
\begingroup
\renewcommand*{\arraystretch}{1.2}
\begin{subequations}\label{eq:macro2D}
\begin{align}
 \begin{pmatrix}
 n \\ \rho u_i \\ \Pi_{ij}
 \end{pmatrix} &=\int dp_xdp_y
 \begin{pmatrix}
 1 \\ p_i \\ \xi_i\xi_j/m
 \end{pmatrix} \phi_{\text{2D}},\\
 \begin{pmatrix}
 \frac{3}{2}n k_B T \\ q_i
 \end{pmatrix} &=\int dp_xdp_y
 \begin{pmatrix}
 1 \\ \xi_i/m
 \end{pmatrix} \left(\frac{\xi_j\xi_j}{2m}\phi_{\text{2D}}+\frac{1}{2}\theta_{\text{2D}}\right)
\end{align}
\end{subequations}
\endgroup
where the sum over repeated Cartesian indices $j \in \{x,\,y\}$ is implicitely understood. In this case, the evolution equations for the reduced distribution functions are:
\begingroup
\renewcommand*{\arraystretch}{1.25}
\begin{equation}
 \frac{\partial}{\partial t}
 \begin{pmatrix}
 \phi_{\text{2D}}\\ \theta_{\text{2D}}
 \end{pmatrix}
+ \frac{p_x}{m}\frac{\partial}{\partial x}
\begin{pmatrix}
 \phi_{\text{2D}}\\ \theta_{\text{2D}}
 \end{pmatrix}
+ F_y\frac{\partial}{\partial p_y}
\begin{pmatrix}
 \phi_{\text{2D}}\\ \theta_{\text{2D}}
 \end{pmatrix}
 = -\frac{1}{\tau} \begin{pmatrix}
 \phi_{\text{2D}}-\phi_{\text{2D}}^S\\ \theta_{\text{2D}}-\theta_{\text{2D}}^S
 \end{pmatrix}
 + \begin{pmatrix}
 J_{1}^{\phi_{\text{2D}}}\\ J_{1}^{\theta_{\text{2D}}}
 \end{pmatrix}\label{eq:evolution_2D}
\end{equation}
\endgroup
where, for brevity, we included only the body force term necessary for the Poiseuille flow.

In the evolution equations above, $\phi_{\text{2D}}^S$ and $\theta_{\text{2D}}^S$ are given by:
\begin{subequations}\label{eq:f2D_expanded}
\begin{align}
\phi_{\text{2D}}^S=f_{2D}^{\text{\tiny MB}}\left[1 + \frac{1-\text{Pr}}{5P_i m k_B T}\left( \frac{\xi_x^2+\xi_y^2}{m k_B T}-4 \right)(\xi_x q_x +\xi_y q_y) \right],\\
\theta_{\text{2D}}^S = k_B T f_{2D}^{\text{\tiny MB}}\left[1 + \frac{1-\text{Pr}}{5P_i m k_B T}\left( \frac{\xi_x^2+\xi_y^2}{m k_BT}-2 \right)(\xi_x q_x+\xi_y q_y) \right]
\end{align}
\end{subequations}
where
\begin{equation}\label{eq:f_mb_xy}
 f_{2D}^{\text{\tiny MB}}=\frac{n}{(2 m \pi k_B T)}\exp{\left(-\frac{\xi_x^2+\xi_y^2}{2 m k_B T}\right).}
\end{equation}
The first order corrections $J_1^{\phi_{2D}} $ and $J_1^{\theta_{2D}}$ in Eqs.~\eqref{eq:evolution_2D} are:
\begin{subequations}\label{eq:j1_expanded_2D}
 \begin{multline}
 J_{1}^{\phi_{\text{2D}}}=-\left[ \xi_x\partial_x\ln\chi + 2\xi_x\partial_x\ln\rho +
 \frac{2}{5}\left(\frac{\xi_x^2}{m k_B T} +\frac{\xi_x^2+\xi_y^2}{2m k_B T} -2 \right)\partial_x u_x \right.  \\ \left.
 +\frac{3\xi_x}{10m}\left(\frac{\xi_x^2+\xi_y^2}{m k_B T}-\frac{2}{3} \right)\partial_x\ln T
 \right]f_{2D}^{\text{\tiny MB}}b\rho\chi
 \end{multline}
\begin{multline}
 J_{1}^{\theta_{\text{2D}}}=-\left[ \xi_x\partial_x\ln\chi + 2\xi_x\partial_x\ln\rho + \frac{2}{5}\left(\frac{\xi_x^2}{m k_B T} +\frac{\xi_x^2+\xi_y^2}{2m k_B T} -1 \right)\partial_x u_x \right. \\ \left.+\frac{3\xi_x}{10m}\left(\frac{\xi_x^2+\xi_y^2}{m k_B T}+\frac{4}{3} \right)\partial_x\ln T
 \right]2 m k_B T f_{2D}^{\text{\tiny MB}}b\rho\chi
\end{multline}
\end{subequations}

\section{Finite-difference Enskog Lattice Boltzmann model}\label{sec:fdlb}

In this section, we outline the FDLB algorithm used to solve Eqs.~\eqref{eq:evolution_1D} and \eqref{eq:evolution_2D} numerically. This algorithm involves two main stages. The first one is the discretization of the momentum space and the second one is the choice of the numerical scheme used to handle the advection term in the evolution equations.

\subsection{Discretization of the momentum space}\label{sec:LB:mspace}

In our work, we use Gauss-Hermite quadrature methods of various orders for the discretization of the momentum space~\cite{H87, SYC06, AS16a, AS16b, AS18, AS2019}. Unlike Ref.~\cite{B23}, where the full-range Gauss-Hermite quadrature was adequate for the use in the Enskog FDLB model for the investigation of problems involving only periodic boundary conditions, namely the propagation of the sound and the shock waves in a one-dimensional domain, in this paper we investigate fluid flow problems bounded by two parallel walls perpendicular to the $x$ axis of the Cartesian system. As known in the kinetic theory of gases, the presence of the walls induces a discontinuity of the distribution function, which becomes more effective at higher values of the Knudsen number. In such cases, the half-range Gauss-Hermite quadrature method in confined fluid flow was already proved to be more accurate and more efficient to capture the effects of the wall-induced discontinuity, when compared to the corresponding full-range quadrature method~\cite{AS16a, AS16b, AS18, AS2019}.

After discretization of the momentum space, the integrals in Eqs.\eqref{eq:macro1D} and \eqref{eq:macro2D} are replaced with sums over the $\mathcal{K}$ elements of the discrete momentum set, i.e. the momentum vectors ${\mathbf{p}}_\kappa$ or the corresponding peculiar momenta ${\boldsymbol{\xi}}_\kappa$, $1 \leq \kappa \leq \mathcal{K}$:
\begingroup
\renewcommand*{\arraystretch}{1.25}
\begin{subequations}\label{eq:macro_LB}
\begin{align}
 \begin{pmatrix}
 n \\ \rho u_{i} \\ \Pi_{i,j}
 \end{pmatrix} &=\sum_{\kappa = 1}^{\mathcal{K}}
 \begin{pmatrix}
 1 \\ p_{\kappa;i} \\ \frac{\xi_{\kappa;j}\xi_{\kappa;j}}{m}
 \end{pmatrix} \phi_{\gamma}({\mathbf{p}}_{\kappa})\, ,\\
 \begin{pmatrix}
 \frac{3}{2} n k_B T \\ q_i
 \end{pmatrix} &=\sum_{\kappa = 1}^{\mathcal{K}} d p_{\kappa;i}
 \begin{pmatrix}
 1 \\ \frac{\xi_{\kappa;i}}{m}
 \end{pmatrix} \left(\frac{\xi_{\kappa;j}\xi_{\kappa;j}}{2m}\phi_{\gamma}({\mathbf{p}}_{\kappa}) + \frac{1}{2}\theta_{\gamma}({\mathbf{p}}_{\kappa}) \right)
\end{align}
\end{subequations}
\endgroup
The indices $i,\, j\,\in {\{x,\,y\}}$ denote the Cartesian components of the vectors $\mathbf{p_{\kappa}}$ and $\boldsymbol{\xi_{\kappa}}$, while $\gamma\in\{\text{1D,\,2D}\}$.

In order to take advantage of the geometry of the channel flows considered in this paper, when investigating the Couette and the Poiseuille flow we solve the evolution equations of the reduced distribution functions by employing the mixed quadratures concept, according to which the quadrature is controlled separately on each axis~\cite{AS16a,BA19,ASS20}. The details regarding the 2D LB model with mixed quadratures are found in Subsec.~\ref{sec:scheme:mixed} below.

\subsubsection{Mixed quadrature LB models with
reduced distribution functions}\label{sec:scheme:mixed}

In mixed quadrature lattice Boltzmann (LB) models, the momentum space is constructed using a direct product rule. This allows for independent construction of quadratures on each axis, by taking into account the flow characteristics.

In the Couette and Poiseuille flows considered in this paper, the presence of diffuse-reflective walls introduces a significant discontinuity in the distribution functions $\phi_{2D}$ and $\theta_{2D}$ when the Knudsen number (${ Kn}$) becomes sufficiently large. For such values of $Kn$, the efficiency of the full-range Gauss-Hermite quadrature on the axis perpendicular to the wall diminishes compared to the half-range Gauss-Hermite quadrature~\cite{AS16b,AS16a,BA19,ASS20}. On the other hand, for axes not bounded by walls, a relatively low-order quadrature (usually a full-range Gauss-Hermite quadrature) is adequate. The detailed information about the full-range Hermite polynomials is provided in Appendix \ref{appendix:fullrange}, which also includes a comparison of simulation results obtained using both full and half range approaches.

When using reduced distribution functions to address 3D flow problems that are homogeneous along the $z$ axis and bounded by walls perpendicular to the $x$ axis, as encountered in the subsequent sections, appropriate LB models employ a half-range Gauss-Hermite quadrature of order $Q_x^h$ along the $x$ axis and a full-range Gauss-Hermite quadrature of order $Q_y$ along the $y$ axis.
The number of momentum vectors used in the FDLB model is ${\mathcal{K}} = \mathcal{Q}_{x} \times \mathcal{Q}_{y}$, where $\mathcal{Q}_x=2Q_x^h$ and $\mathcal{Q}_{y}=Q_y$. This technique allows one to reduce the computational costs~\cite{AS16b, AS18,ASS20}. Note that, when simulating the Fourier flow there is no variation of the reduced distribution functions along the $y$ axis. In this case, the sole use of the half-range Gauss-Hermite quadrature on the $x$ axis is sufficient and the mixed quadratures are no longer needed.

\subsubsection{Half-range Gauss-Hermite quadrature}\label{sec:LB:hh}

To ensure the recovery of the half-range moments $M_s^\pm$ 
and $M_s^{(\mathrm{eq}),\pm}$ of finite order $s$ for the distribution functions $f$, half-range quadratures are employed on the $x$ axis. The moments are defined as follows:
\begin{equation}
 \begin{pmatrix}
 M_s^+ \\
 M_s^{{( \mathrm{eq} )}, +} \rule{0mm}{7mm}
 \end{pmatrix} = \int_0^\infty dp\,
 \begin{pmatrix}
 f(p) \\
 f^{\mathrm{eq}}(p) \rule{0mm}{7mm}
 \end{pmatrix} p^s, \qquad
 \begin{pmatrix}
 M_s^- \\
 M_s^{{( \mathrm{eq} )}, -} \rule{0mm}{7mm}
 \end{pmatrix} = \int_{-\infty}^0 dp\,
 \begin{pmatrix}
 f(p) \\
 f^{\mathrm{eq}}(p) \rule{0mm}{7mm}
 \end{pmatrix} p^s.\label{eq:momspm}
\end{equation}

% To achieve the recovery of these half-range integrals,
The half-range Gauss-Hermite quadrature on the $x$ axis is defined with respect to the weight function $\omega(p)$:
\begin{subequations}
\begin{eqnarray}
 \omega(p_x) &=& \frac{1}{\sqrt{2\pi}} e^{\bar{p}_x^2/2},\quad  \bar{p}_x\equiv p_x / p_0 \label{eq:hh_quad_weight}\\
 \int_0^\infty dp\, \omega(\bar{p}_x) P_s(\bar{p}_x) &\simeq& \sum_{k_x = 1}^{Q_{x}^{h}} w(p_{k_x}) P_s(p_{k_x}),\label{eq:hh_quad}
\end{eqnarray}
\end{subequations}
where $P_s$ represents a polynomial of order $s$, $p_0$ is a characteristic momentum scale and the equality \eqref{eq:hh_quad} is exact if the number of quadrature points $Q_x^h$ satisfies $2Q_x^h > s$~\cite{H87,SB15}. In this paper we set $p_0=\sqrt{m_{ref}k_BT_{ref}}$.
The quadrature points $p_{k_x}$ ($k_x=1,2,\dots,Q_x^h$) are the positive roots of the half-range Hermite polynomial $\hh_Q(p)$ of order Q. The quadrature weights $w_{k_x}^h$ are given by~\cite{H87,SB15,AS16b,AS16a}:
\begin{equation}
 w_{k_{x}}^h =
  \frac{p_{k_x} a_Q^2}{\hh_{Q+1}^2(p_{k_x}) \left[p_{k_x} + \hh_{Q}^2(0)/\sqrt{2\pi}\right]},
 \label{eq:hh_wk}
\end{equation}
where $a_Q = \hh_{Q+1,Q+1} / \hh_{Q,Q}$ and $\hh_{\ell,s}$ represents the coefficient of $p^s$ in $\hh_\ell(p)$:
\begin{equation}
 \hh_\ell(p) = \sum_{s = 0}^\ell \hh_{\ell, s} \bar{p}^s.\label{eq:hh_exp}
\end{equation}
The half-range Hermite polynomials are normalized according to:
\begin{equation}
 \int_0^\infty dp\, \omega(p) \hh_\ell(p) \hh_{\ell'}(p) = \delta_{\ell,\ell'}.
 \label{eq:hh_ortho}
\end{equation}

To apply the half-range Gauss-Hermite quadrature, the distribution function $f$ and the equilibrium distribution function $f^{\mathrm{eq}}$ must be expanded using the half-range Hermite polynomials. Since these polynomials are defined only on half of the momentum axis, $f$ is split using the Heaviside step function $\theta(p)$:
\begin{equation}
 f(p) = \theta(p) f^+(p) + \theta(-p) f^-(p), \qquad
 \theta(p) =
 \begin{cases}
  1,& p> 0,\\
  0, & p < 0.
 \end{cases}
\end{equation}
The functions $f^+(p)$ and $f^-(p)$ are defined on the positive and negative momentum semi-axis, respectively, and can be expanded as:
\begin{equation}
 f^+ = \frac{\omega(p)}{p_0} \sum_{\ell = 0}^\infty \mathcal{F}^+ \hh_\ell(p), \qquad
 f^- = \frac{\omega(-p)}{p_0} \sum_{\ell = 0}^\infty \mathcal{F}^- \hh_\ell(-p).
 \label{eq:hh_f}
\end{equation}
The coefficients $\mathcal{F}^\pm$ can be obtained using the orthogonality of the half-range Hermite polynomials:
\begin{equation}
 \mathcal{F}^+_\ell = \int_0^\infty dp\, f(p) \hh_\ell(p), \qquad
 \mathcal{F}^-_\ell = \int_{-\infty}^0 dp\, f(p) \hh_\ell(-p).
 \label{eq:hh_F}
\end{equation}

Expanding $f^{\mathrm{eq}}$ in a similar manner, with expansion coefficients $\mathcal{G}^\pm_\ell$, the half-range moments can be recovered:
\begin{equation}
 \begin{pmatrix}
 M_s^+\\
 M_s^- \rule{0mm}{7mm}
 \end{pmatrix}
 = \sum_{\ell = 0}^\infty \frac{1}{\ell!}
 \begin{pmatrix}
 \mathcal{F}_\ell^+\\
 \mathcal{F}_\ell^- \rule{0mm}{7mm}
 \end{pmatrix}
 \int_{0}^\infty dp\,\omega(p)\, \hh_\ell(p) (\pm p)^s.\label{eq:momspm_aux}
\end{equation}
Truncating the expansion at $\ell = Q-1$ ensures that a quadrature of order $Q$ can recover the moments for $0 \leq s \leq Q$. The discrete momentum set of the one-dimensional half-range Gauss-Hermite lattice Boltzmann model consists of $\mathcal{Q} = 2Q$ elements, defined as:
\begin{equation}
 \bar{p}_{k_x} = p_0 p_{k_x}, \qquad \bar{p}_{k_x + Q} = -p_{k_x} \qquad
 (1 \le k \le Q_x^h).\label{eq:hh_pk}
\end{equation}
Thus, the half-range moments are recovered as:
\begin{equation}
 M_s^+ = \sum_{k = 1}^Q f_k p_{k_x}^s, \qquad
 M_s^- =\sum_{k = Q+1}^{2Q} f_k p_{k_x}^s,
\end{equation}
where
\begin{equation}
 f_k = \frac{w_k^{h} p_0}{\omega(p_k)} f(p_k), \qquad
 f_{k + Q} = \frac{w_k^{h} p_0}{\omega(p_k)} f(-p_k) \qquad
 (1 \le k \le Q_x^h).\label{eq:hh_fk}
\end{equation}

Now let us consider the expansion of $f^{\mathrm{eq}}$ in terms of the half-range Hermite polynomials. We can write $f_x^{\text{MB}}= \theta(p) g_+(p) + \theta(-p) g_-(p)$, where
\begin{equation}
 g_\pm = \frac{\omega(\abs{p})}{p_0} \sum_{\ell = 0}^\infty \mathcal{G}^\pm_\ell
 \hh_\ell(\abs{p}).\label{eq:hh_gk}
\end{equation}
The expansion coefficients $\mathcal{G}^\pm_\ell$ can be obtained in a similar way to Eq.~\eqref{eq:hh_F}.

For more information on one-dimensional lattice Boltzmann models based on half-range Gauss-Hermite quadratures, we refer the reader to Refs.~\cite{AS16a, BA19, ASS20}.

\subsection{Numerical schemes}\label{sec:LB:scheme}

For reader's convenience, in this subsection we detail the numerical schemes employed to solve the evolution equations \eqref{eq:evolution_1D} and \eqref{eq:evolution_2D}, namely the third-order total variation diminishing (TVD) Runge-Kutta method for time-stepping, the fifth-order WENO-5 advection scheme, and the $6$th order central difference scheme used for gradient evaluation.

Diffuse boundary conditions are applied on both walls/boundaries. This implies that the molecules striking the walls are re-emitted according to the Maxwellian distribution with parameters $T_w$ and $U_w$, where $T_w$ and $U_w$ represent the predetermined wall temperature and velocity, respectively. The value of $n_w$ is determined by satisfying the impenetrable condition~\cite{AS16a, AK02}.

\subsubsection{Third-order TVD Runge-Kutta method}

After the momentum space discretization, it is convenient to cast the Enskog equation \eqref{eq:enskog_approx} in the following form:
\begin{equation}\label{eq:cast}
 \partial_t f_{\kappa} = L[f_{\kappa}], \qquad
 L[f_{\kappa}] = - \frac{{p}_{\kappa;\,x}}{\,m\,}\cdot \nabla_x f_{\kappa} + (\partial_{p_y} f)_\kappa
-\frac{1}{\tau}[f_{{\kappa}} - f^{S}_{{\kappa}}]+J_{1,\kappa}.
\end{equation}
in order to implement the time-stepping algorithm. In the above the subscript $\kappa$ refers to the discretized functions corresponding to $\xi_\kappa$.

The third-order Runge-Kutta integrator gives the following three-step algorithm for computing the values of $f_{\kappa}$ at time $t+ \delta t$~\cite{SO88,GS98,RZ13}:
\begin{align}
 f_{\kappa}^{(1)}(t) =& f_{\kappa}(t) + \delta t \, L[f_{\kappa}(t)], \nonumber\\
 f_{\kappa}^{(2)}(t) =& \frac{3}{4} f_{\kappa}(t) + \frac{1}{4} f_{\kappa}^{(1)}(t) +
 \frac{1}{4} \delta t\, L[f_{\kappa}^{(1)}(t)],\nonumber\\
 f_{\kappa}(t+\delta t)
 =& \frac{1}{3} f_{\kappa}(t) + \frac{2}{3} f_{\kappa}^{(2)}(t) +
 \frac{2}{3} \delta t\, L[f_{\kappa}^{(2)}(t)]. \label{eq:rk3}
\end{align}

\subsubsection{WENO-5 advection scheme}

The advection term which appears in Eq.~\eqref{eq:cast} above, namely $p_{\kappa;x}\cdot \nabla_x f_{\kappa}/m$ is computed using the Weighted Essentially Non-Oscillatory scheme of order $5$ (WENO-5) along each coordinate~\cite{GXZL11,JS96}. We will describe in the following the one-dimensional case. Assuming that the flow domain is discretized using $1 \le i \le N$ nodes on the $x$ axis, the advection term becomes:
\begin{equation}
 \left(\frac{p_{\kappa}}{m}  \partial_x f_{\kappa}\right)_{\kappa; i} =
 \frac{\mathcal{F}_{\kappa; i + 1/2} - \mathcal{F}_{k; i - 1/2}}
 {\delta s}
\end{equation}
where $\mathcal{F}_{k;i+1/2}$ represents the flux of $f$ advected with velocity $p_{\kappa} / m$ through the interface between the cells centered on $\bm{x}_{i}$ and $\bm{x}_{i+1}$. The construction of these fluxes is summarized below, under the assumption of a positive advection velocity $p_{\kappa} / m > 0$. In this case, the flux $\mathcal{F}_{k; i+1/2}$ can be computed using the following expression~\cite{GXZL11}:
\begin{equation}\label{eq:weno5_flux_x}
\mathcal{F}_{\kappa,i+1/2} = \overline{\omega}_1\mathcal{F}^1_{\kappa,i+1/2} +
\overline{\omega}_2\mathcal{F}^2_{\kappa,i+1/2} + \overline{\omega}_3\mathcal{F}^3_{\kappa,i+1/2}.
\end{equation}

The interpolating functions $\mathcal{F}^q_{\kappa,i + 1/2}$ ($q = 1,2,3$) are given by:
\begin{align}
\mathcal{F}^1_{\kappa,i+1/2} =& \frac{p_{\kappa}}{m} \left(\frac{1}{3}f_{\kappa,i-2} - \frac{7}{6} f_{\kappa,i-1} + \frac{11}{6} f_{\kappa,i}\right), \nonumber \\
\mathcal{F}^2_{\kappa,i+1/2} =& \frac{p_{\kappa}}{m} \left(-\frac{1}{6}f_{\kappa,i-1} + \frac{5}{6} f_{\kappa,i} + \frac{1}{3} f_{\kappa,i+1}\right), \nonumber \\
\mathcal{F}^3_{\kappa,i+1/2} =& \frac{p_{\kappa}}{m} \left(\frac{1}{3}f_{\kappa,i} + \frac{5}{6} f_{\kappa,i+1} - \frac{1}{6} f_{\kappa,i+2}\right).
\end{align}

The weighting factors $\overline{\omega}_q$ appearing in Eq.~\eqref{eq:weno5_flux_x} are given by:
\begin{equation}
\overline{\omega}_q = \frac{\widetilde{\omega}_q}{\widetilde{\omega}_1+\widetilde{\omega}_2+\widetilde{\omega}_3}, \qquad
\widetilde{\omega}_q = \frac{\delta_q}{\varphi^2_q}.
\label{eq:weno5_omega_x}
\end{equation}

The ideal weights $\delta_q$ are:
$\delta_1 = \frac{1}{10}, \, \delta_2 = \frac{6}{10},\, \delta_3 = \frac{3}{10},$
while the smoothness indicators $\varphi_q$ can be computed as follows:
\begin{align}
\varphi_1 =& \frac{13}{12} \left(f_{i-2} -2f_{i-1} + f_i \right)^2
+ \frac{1}{4} \left( f_{i-2} - 4f_{i-1} + 3f_i \right)^2,
\nonumber \\
\varphi_2 =& \frac{13}{12} \left(f_{i-1} -2f_{i} + f_{i+1} \right)^2
+ \frac{1}{4} \left( f_{i-1} - f_{i+1} \right)^2,
\nonumber \\
\varphi_3 =& \frac{13}{12} \left(f_{i} -2f_{i+1} + f_{i+2} \right)^2
+ \frac{1}{4} \left( 3f_{i} -4 f_{i+1} + f_{i+2} \right)^2.
\label{eq:weno5_sigma}
\end{align}
where the index $\kappa$ was omitted for brevity.
The computation of the weighting factors $\overline{\omega}_q$ \eqref{eq:weno5_omega_x} implies the division between the ideal weights $\delta_q$ and the smoothness indicators $\varphi_q$ \eqref{eq:weno5_sigma}.
To avoid division by $0$ when either one, two, or all three of the indicators of smoothness vanish, we follow Refs.~\cite{BA18,BA19} and compute the weighting factors $\overline{\omega}_q$ directly in the limiting cases when any of the smoothness indicators vanishes.

\subsubsection{\label{sec:gardients} The 6th order central difference scheme for the gradient operator}

For evaluating the gradients we employ the $6$th order central difference scheme~\cite{F88}:
\begin{multline}
 \partial_x Q(x)=
 \frac{1}{\Delta x}\left[-\frac{1}{60} Q(x-3\Delta x)+\frac{3}{20} Q(x-2\Delta x)-\frac{3}{4} Q(x-\Delta x)\right.\\
 \left.+\frac{3}{4} Q(x+\Delta x)-\frac{3}{20} Q(x+2\Delta x)+\frac{1}{60} Q(x+3\Delta x)\right]
\end{multline}
where $Q\in\{\ln \rho,u,\ln T \}$.

\section{\label{sec:results} Numerical results}

Our attention is directed towards the study of dense gas flows confined between two infinite parallel plates which are perpendicular to the $x $ direction. The plates are located at $x=-L/2$ and $L/2$, respectively, while the diffuse reflection boundary conditions are applied at $x=\pm(L-\sigma)/2$. The confinement ratio is defined as $R=L/\sigma$, where $L$ is the physical domain width and $L_c=L-\sigma$ is the width of the computational domain. The FDLB results are compared to the simulation results obtained with the particle method (DSMC) which is briefly presented in Appendix \ref{appendix:pm}. If not stated otherwise, the time step was set to $\Delta t=10^{-3}$ and the lattice spacing (cell length for particle method) at $\Delta x=\sigma/100$. A number of $1000$ particles per cell was used in the particle method in order to obtain smooth profiles of macroscopic quantities.

For our simulations, we have chosen the following value sets of the confinement ratio $R$ and the mean (initial) reduced density $\eta_0$, namely $R\in\{4,\,10\}$ and $\eta_0\in\{0.01,\,0.10,\,0.20\}$. We have chosen these values to encompass a wide range of Knudsen numbers, spanning from the slip to the early transition regime. The objective is to specifically emphasize the unique characteristics of fluid flow when dense gas effects and confinement are involved.

In the following subsections, we examine various scenarios to assess the effectiveness of the proposed model in recovering the flow of a dense gas bounded by parallel plates.
When not stated otherwise, the orders $Q_x^h$ of the half-range quadratures used in this study for the three flow problems (Fourier, Couette, and Poiseuille) are listed in Table \ref{tab:quad}. These values were obtained by performing a convergence test, for which the following error evaluation was used~\cite{AS16a}:
\begin{equation}
 \epsilon_M^{\delta x}=\frac{\max\left[ M(x) -M_{ref}(x) \right]}{\max[M_{ref}(x)]- \min[M_{ref}(x)] }
\end{equation}
where $M\in\{T,u_y,u_y\}$, for the Fourier, Couette, and Poiseuille flow, respectively, and $\delta x$ is the lattice spacing which is kept fixed for the convergence test. The reference profile $M_{ref}$ is obtained using $Q_x^h=200$ quadrature points. The convergence test is satisfied when $\epsilon_m^{\delta x=\sigma/100}<0.01$, which represents a $1\%$ error with respect to the macroscopic quantity maximum variation in the channel. In the case of the 2D flows a quadrature of order $Q_y=5$ was imposed in the longitudinal direction for both reduced distributions~\cite{ASS20}.

The typical runtime for a PM simulation is about $3.5\times 10^3$ s for $\eta_0=0.01$ and $1.85\times 10^4$ s for $\eta_0=0.2$, irrespective of the flow studied, while the FDLB, using a quadrature order of $Q_x^h=8$, takes around 17s for the 1D case of Fourier flow, and 100s for the 2D case of Couette and Poisseuille flow. The running time for FDLB is independent of the reduced density employed but it is directly proportional to the quadrature order, resulting in a minimum runtime ratio of $\approx150$ in the Fourier flow case ($Q_x^h=11$ for $\eta_0=0.01$), and a minimum runtime ratio $\approx10$ in the Poisseuille case with $Q_x^h=29$ at $eta_0=0.01$. These times were recorded using $R=4$ and a single core of an Intel(R) Xeon(R) Gold 6330 CPU running at 2.0GHz.

\begin{table}
	\begin{center}
		\begin{tabular*}{\columnwidth}{@{\extracolsep{\stretch{1}}}*{7}{l||cc|cc|cc}@{}}
			\toprule
			  & Fourier & & Couette & & Poiseuille \\ \hline
    $\eta_0\backslash R $  & 4 & 10 & 4 & 10 & 4 & 10 \\ \hline\hline
			 0.01 & 11 & 8 & 8 & 8 & 29 & 15 \\ \hline
			 0.1 & 8 & 8 & 8 & 8 & 8 & 8 \\ \hline
    0.2 & 8 & 8 & 8 & 8 & 8 & 8\\ \hline
			 \hline
		\end{tabular*}
	\end{center}
	\caption{ Quadrature order $Q^h_x$ for the parameters used in this study. }
	\label{tab:quad}
\end{table}

\subsection{Dense gas at rest near a reflective wall }\label{sec:gas_at_rest}

\begin{figure*}
\includegraphics[width=0.49\linewidth]{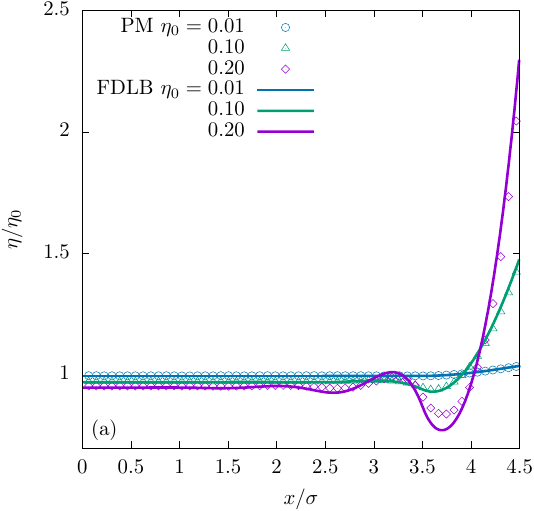}
\includegraphics[width=0.49\linewidth]{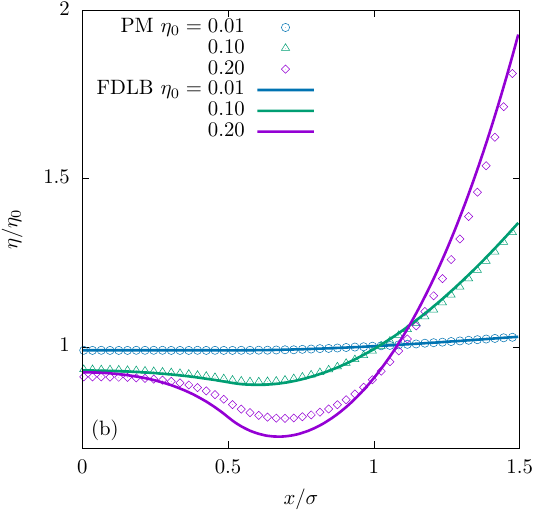}
\caption{\label{fig:dens_stationary} Dense gas at rest near a reflective wall: normnalized density profiles $\eta / \eta_0$ for three values of the mean reduced density $\eta_0\in\{0.01,\,0.10,\,0.20\}$ and two values of the confinement ratios $R=10$ (a) and $R=4$ (b).}
\end{figure*}

To begin, we consider a hard-sphere gas confined between two stationary parallel walls, kept at the temperature $T_w = 1$. We first assess the accuracy of the proposed model in reproducing the reduced density profile near a wall. Due to the symmetry of the simulated problem, only the half-channel plot is presented ($x\in[0:L_c]$), as it will be also the case in the Couette and the Poiseuille flows to be discussed further.
In Fig. \ref{fig:dens_stationary} one may observe that the stationary profile of the reduced density $\eta$ is non-monotonic near the wall, unlike in the case of a dilute gas. Indeed, it is important to consider that when a fluid particle is located at a distance less than $\sigma$ from the wall, a portion of its surface remains protected from collisions. This occurs because there is insufficient space available for a second molecule to occupy that region. As a result, the particle is pushed toward the wall. It is important to note that an oscillating density profile near the wall is a characteristic feature of dense gases. These density variations emerge within a region approximately equivalent to the molecular diameter, and their intensity diminishes as the $\eta_0\rightarrow 0$. Consequently, in the Boltzmann limit, the density becomes spatially uniform.
To assess the accuracy of the FDLB model, we compare the density profiles to the profiles obtained using the particle method (PM). As expected, the difference between the FDLB and the PM profiles becomes significantly larger when the mean reduced density $\eta_0$ increases. As seen in Fig.\ref{fig:dens_stationary}, a fairly good agreement between the PM and the FDLB results is obtained up to the mean value $\eta_0=0.1$. For larger values of $\eta_0$, substantial differences can be observed regardless of the value of the confinement ratio $R$. This value $\eta_0$ can be assumed to be the limit of the simplified Enskog collision model, for a reasonable accuracy.

\subsection{\label{sec:fourier} Fourier flow}

In this subsection, we will analyze the thermal transfer in a dense gas confined between two infinite parallel plates. The left and the right plate temperatures are fixed at $T_L=T_0-\Delta T$ and $T_R=T_0+\Delta T$ respectively. We will explore two values of the temperature difference $\Delta T\in\{0.1,0.5\}$, corresponding to $T_R/T_L\in\{11/9,\,3\}$. For each value of the temperature difference $\Delta T$, the simulations were conducted using three values of the mean reduced density
$\eta_0\in\{0.01,0.1,0.2\}$ and two values of the confinement ratio $R\in\{4,10\}$($L_c\in\{3,9\}$).

Fig.\ref{fig:termtrans_DT01} presents the results for the smaller temperature difference $\Delta T=0.1$. The profiles of the normalized reduced density $\eta/\eta_0$ and the temperature $T$ are shown in the first row and the second row, respectively. Excellent agreement is obtained for both confinement ratios at small values of density $\eta_0$, while at higher values of $\eta_0$ the density and temperature present discrepancies between the two simulation methods, especially near the walls. Nevertheless, these results are in good relative agreement.

Figure \ref{fig:termtrans_DT05} depicts the outcomes for a significantly larger temperature difference $\Delta T=0.5$, corresponding to the wall temperature ratio $T_R/T_L=3$. The density profiles resemble those observed for smaller temperature differences. However, in the case of temperature profiles for high values of the mean reduced density $\eta_0$, noticeable discrepancies are observed also in the bulk of the fluid (the relative error with respect to the PM results is approximatively $5\%$).

\begin{figure*}
\includegraphics[width=0.49\linewidth]{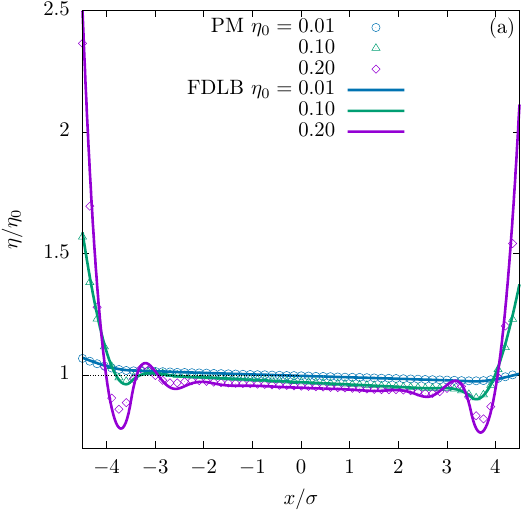}\hfill
\includegraphics[width=0.5\linewidth]{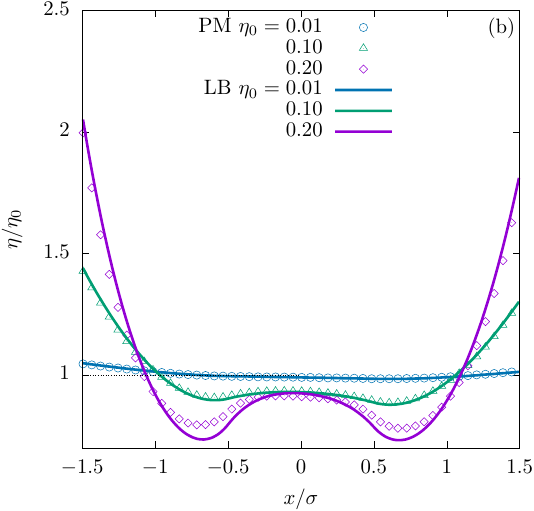}\\
\includegraphics[width=0.49\linewidth]{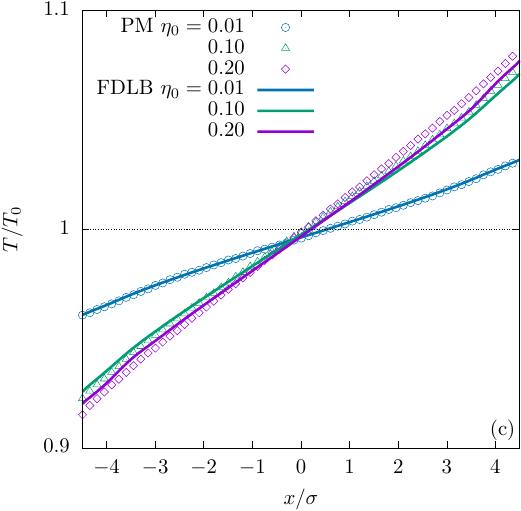}\hfill
\includegraphics[width=0.5\linewidth]{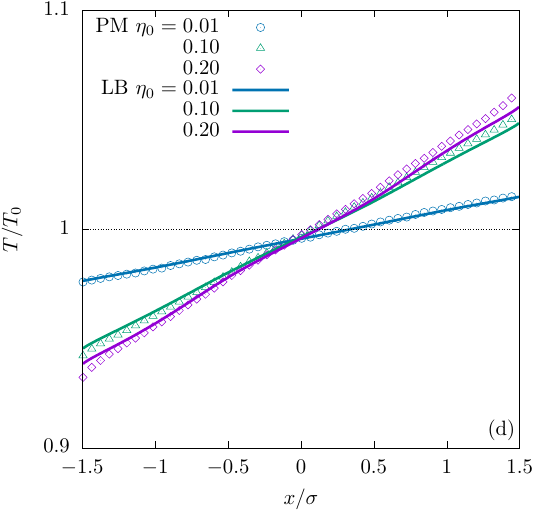}
\caption{\label{fig:termtrans_DT01} Fourier flow:
 Heat transfer at wall temperature difference $\Delta T=0.1$ and three values of the mean reduced density $\eta_0$: Transversal profiles of the ratio $\eta / \eta_0$ (first row) and of the temperature $T$ (second row) for $R=10$ (left column) and $R=4$ (right column).
}
\end{figure*}

\begin{figure*}
\includegraphics[width=0.49\linewidth]{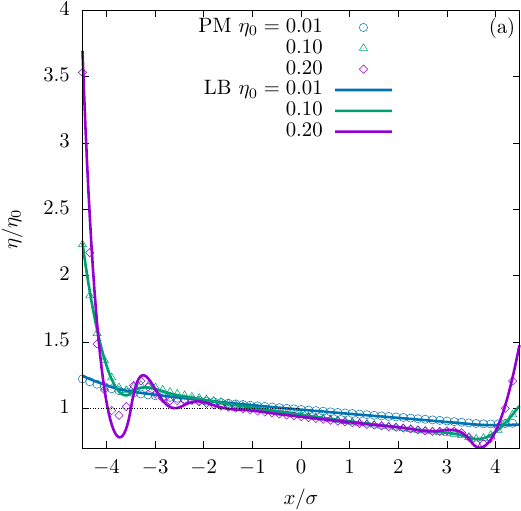}\hfill
\includegraphics[width=0.5\linewidth]{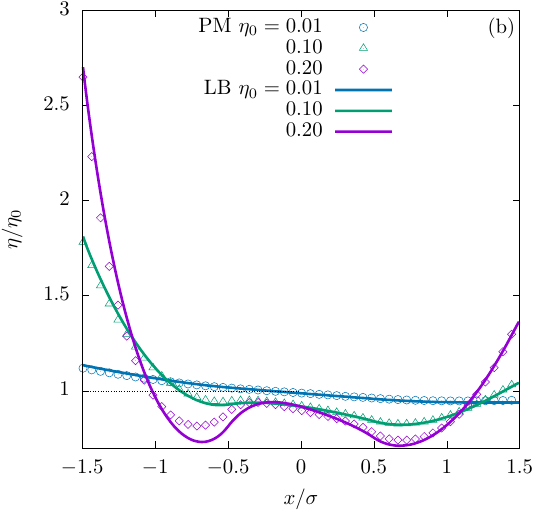}\\
\includegraphics[width=0.49\linewidth]{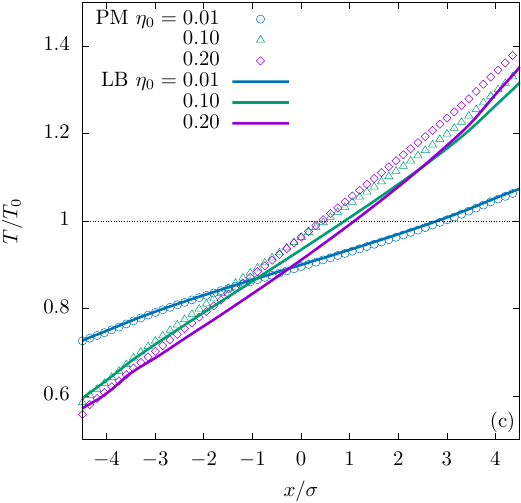}\hfill
\includegraphics[width=0.50\linewidth]{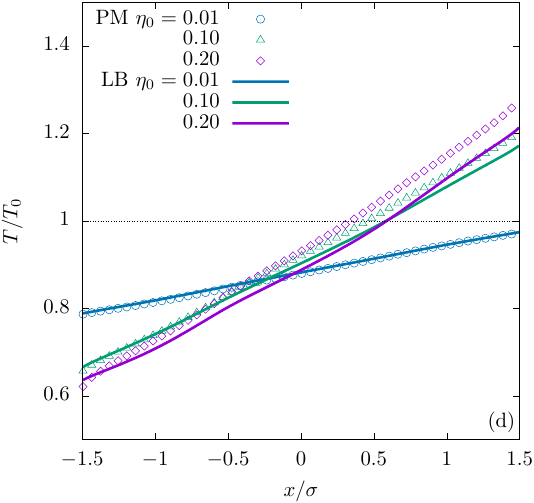}
\caption{\label{fig:termtrans_DT05}Fourier flow: Heat transfer at wall temperature difference $\Delta T=0.5$ and three values of the mean reduced density $\eta_0$ : Transversal profiles of the ratio $\eta / \eta_0$ (first row) and of the temperature $T$ (second row) for $R=10$ (left column) and $R=4$ (right column).}
\end{figure*}

According to the conservation laws, the transversal heat flux $q_x$ is constant through the channel. The profiles of the transversal heat flux $q_x$, obtained with both the FDLB and the PM models are presented in Fig. \ref{fig:termtrans_qx}. The transversal heat flux is plotted in the left panel for $R\in\{4,10\}$ and $\Delta T\in\{0.1,0.5\}$. As the dilute gas limit is approached, the two methods agree perfectly, and the departure of the FDLB results from the PM results for increased values of $\eta_0$ is observed. On the other hand, in the right panel, we plot the ratio $q_x/\eta_0$ with respect to various values of the Knudsen number $Kn$ obtained by varying the width of the computational domain at a constant value of the reduced density $\eta_0$. The monotonic increase of the heat flux with respect to the Knudsen number encountered for the dilute gas~\cite{BCP68,SS98} is no longer present for dense gases at relatively high values of the reduced density $\eta_0$. Indeed, one may observe a local maximum of the heat flux which emerges in both the FDLB and the PM models. This maximum is further investigated using the PM in Appendix~\ref{appendix:heat_flux}, where the two components of the heat flux (kinetic and potential) are evaluated.

\begin{figure*}
\includegraphics[width=0.49\linewidth]{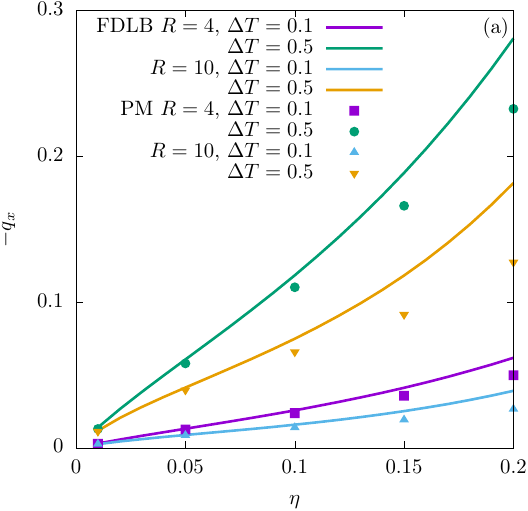}\hfill
\includegraphics[width=0.505\linewidth]{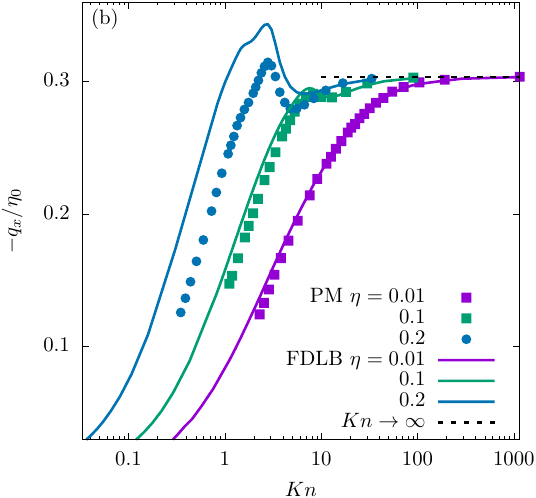}
\caption{\label{fig:termtrans_qx} Fourier flow: (a) Heat flux $q_x$ profiles at temperature difference of $\Delta T=\{0.1,0.5\}$, confinement ratios of $R=\{4,10\}$. (b) Dependence of the ratio $q_x/n_0$ with respect to the Knudsen number $Kn$ for 3 values of the reduced density $\eta\in\{0.01,0.1,0.2\}$.}
\end{figure*}

\subsection{\label{sec:couette} Couette flow}

In this subsection, we will analyze the Couette flow of a gas confined between two infinite parallel plates. The plates move in opposite directions along the $y$ axis with fixed velocity $U_0=\sqrt{k_B T/m}$, i.e. $U_L=-U_0$ and $U_R=U_0$. The simulations were conducted for three values of the mean reduced density, $\eta_0\in\{0.01,0.1,0.2\}$ and two values of the confinement ratio $R\in\{4,10\}$, corresponding to $L_c\in\{3,9\}$.

In Fig.\ref{fig:couette_dens_u01} we plot the transversal density profile on the right half $[0,\, L_c]$ of the computational domain. One can easily recognize the same feature as in the case of stationary dense gas, namely the formation of a layer in the proximity of each wall. The layer width is of the order of $\sigma$ (the molecular diameter). Excellent agreement between the FDLB and the PM profiles is observed up to $\eta_0=0.1$. The corresponding velocity and temperature profiles are presented in Fig. \ref{fig:couette_veltemp_u1}, for the two values of the confinement ratio considered in this paper. Very good agreement between the FDLB and the PM profiles is obtained throughout the considered range of the reduced density $\eta_0$, with small discrepancies observed near the wall when the value of the confinement ratio is large enough ($R=10$). For the smaller value of the confinement ratio ($R=4$), the discrepancies are larger, but this is expected behavior due to the severe approximations involved in the derivation of the simplified Enskog collision operator.

In addition to the transverse component $q_x$ of the heat flux, the Couette flow exhibits a non-zero longitudinal heat flux $q_y$, which is a distinct microfluidics phenomenon. The corresponding results are presented in Figure \ref{fig:couette_qxy_u1}, where the left and the right panels represent the simulation results conducted with the values $R=10$ and $R=4$ of the confinement ratio, respectively. Excellent agreement between the FDLB and the PM results is observed when the reduced density is small enough ($\eta_0=0.01$). For larger values of the reduced density, the discrepancies become noticeable. The plot specifically showcases the right half of the channel, where the transverse heat flux values are positive, and the longitudinal component consistently exhibits negative values across the parameter range. Due to the fact that the longitudinal heat flux, $q_y$, does not arise from a temperature gradient (known as the direct phenomenon), its behavior is expected to depend on higher-order transport coefficients. Hence, it is unsurprising that these cross phenomena are not accurately captured by the model when $\eta_0$ is large enough.

\begin{figure}
\includegraphics[width=0.49\linewidth]{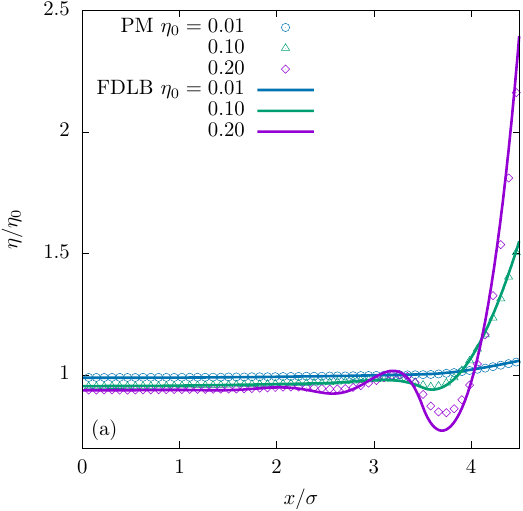}\hfill
\includegraphics[width=0.502\linewidth]{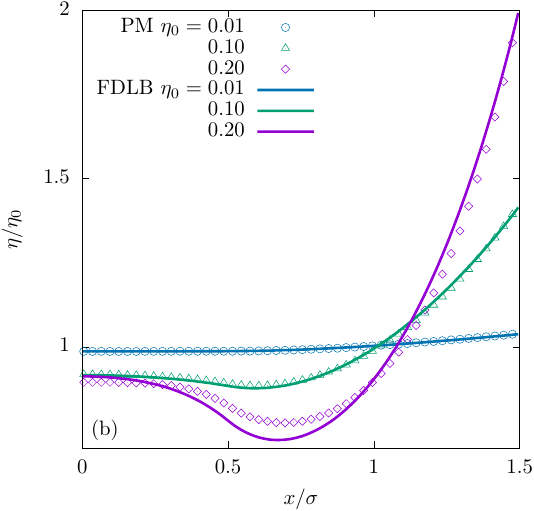}
\caption{\label{fig:couette_dens_u01} Couette flow: Normalised reduced density profile $\eta/\eta_0 $ at a wall velocity of $U_w=1$, three values of the mean reduced density $\eta_0=\{0.01,0.1,0.2\}$ and (a) $R=10$ and (b) $R=4$.}
\end{figure}

\begin{figure}
\includegraphics[width=0.495\linewidth]{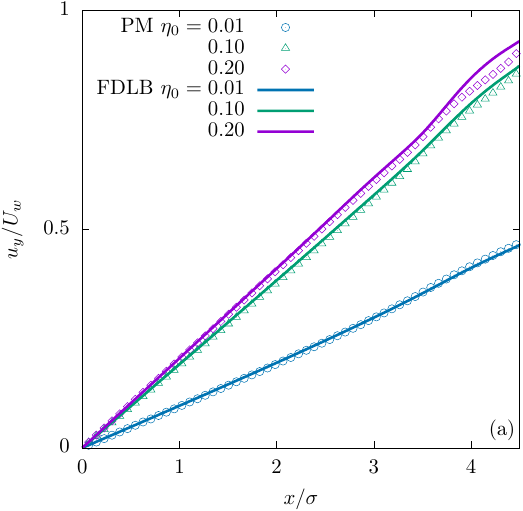}\hfill
\includegraphics[width=0.49\linewidth]{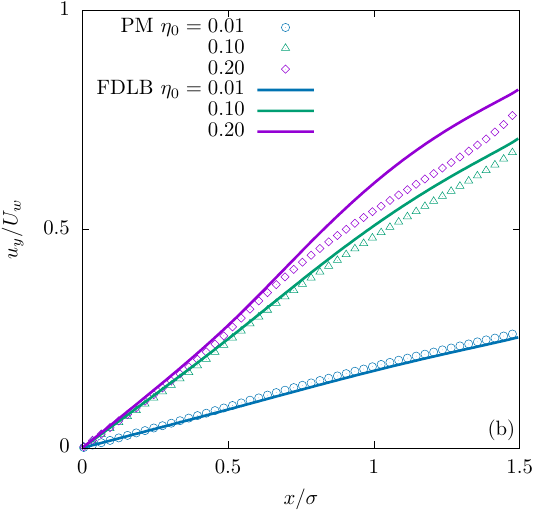}\\
\includegraphics[width=0.495\linewidth]{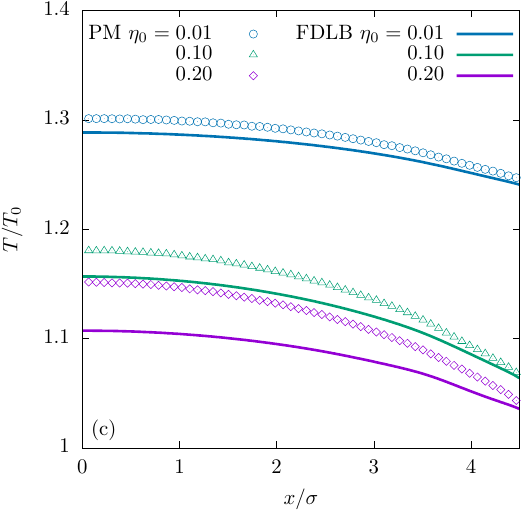}\hfill
\includegraphics[width=0.49\linewidth]{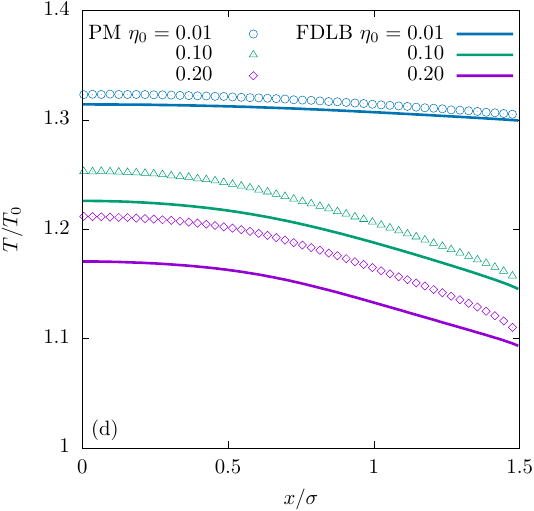}
\caption{\label{fig:couette_veltemp_u1} Couette flow: velocity and temperature profiles at a wall velocity of $U_w=1$, three values of the mean reduced density $\eta_0=\{0.01,0.1,0.2\}$ and $R=10$ (left column) and $R=4$ (right column).}
\end{figure}

\begin{figure}
\includegraphics[width=0.495\linewidth]{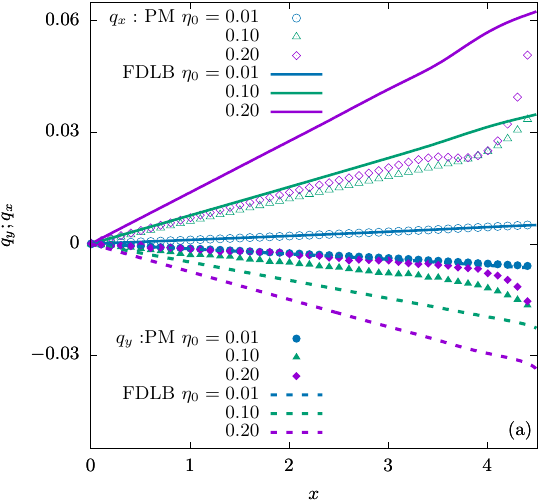}\hfill
\includegraphics[width=0.495\linewidth]{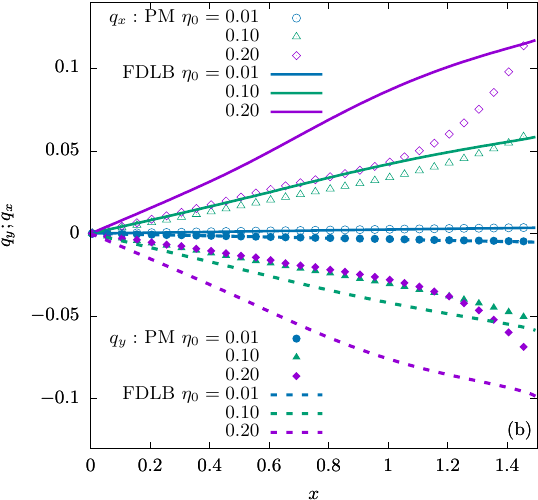}
\caption{\label{fig:couette_qxy_u1} Couette flow: Profiles of both the transversal ($q_x > 0$) and the longitudinal ($q_y<0$) heat fluxes for the wall velocity $U_w=1$, three values of the mean reduced density $\eta_0=\{0.01,0.1,0.2\}$ and two values of the confinement ratio (a) $R=10$ and (b) $R=4$. }
\end{figure}

\subsection{\label{sec:poisseulle} Poiseuille flow}

In this section, we examine the Poiseuille flow, which is generated by an external acceleration $a_y$ acting parallel to the plates. The results are grouped into linear and non-linear flows, based on the driving force magnitude. Nonlinearity arises when the heat generated by viscous dissipation cannot be adequately dissipated through the diffusive boundary conditions. To ensure that the response to the driving force remains within the linear regime, a specific driving force of $a_y=0.001$ has been chosen, as detailed below.

\subsubsection{\label{sec:small_a} Linear regime}

\begin{figure}
\includegraphics[width=0.48\linewidth]{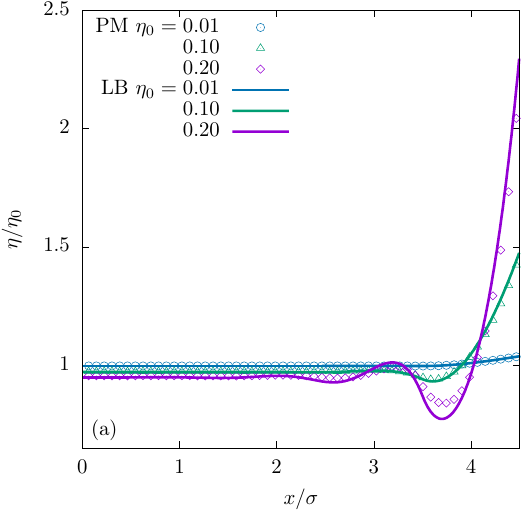}\hfill
\includegraphics[width=0.49\linewidth]{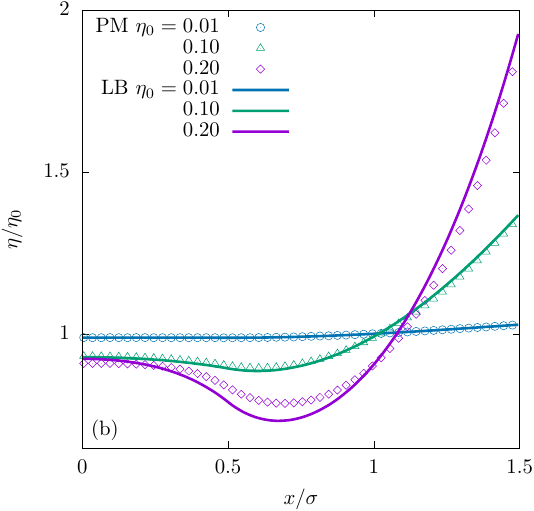}\\
\includegraphics[width=0.495\linewidth]{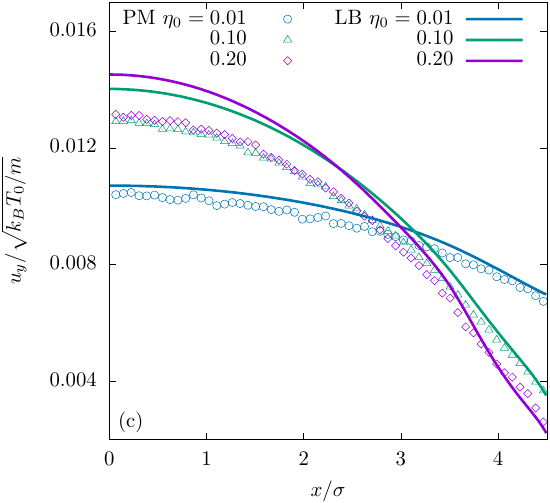}\hfill
\includegraphics[width=0.5\linewidth]{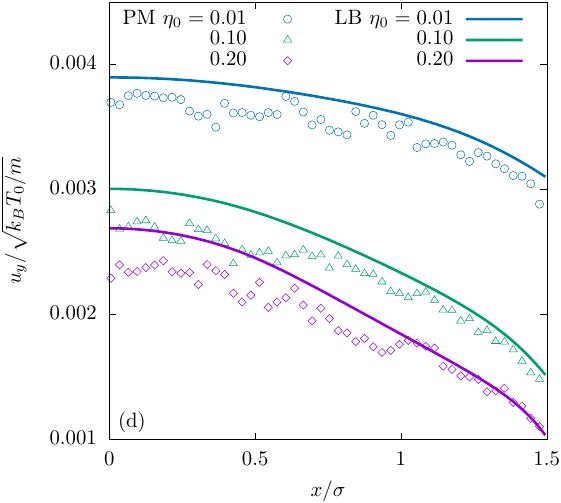}
\caption{\label{fig:poisseuille_dens_a0001} Poiseuille flow (linear regime): Density and velocity at an external acceleration of $a_y=0.001$, three values of the mean reduced density $\eta_0=\{0.01,0.1,0.2\}$ and $R=10$ (left column) and $R=4$ (right column).}
\end{figure}

At first, we will look at a flow with a small acceleration $a_y=0.001$. In this case, the flow deviates slightly from the equilibrium, hence only the velocity field has a statistically significant variation. In Fig. \ref{fig:poisseuille_dens_a0001} we take advantage of the symmetry of the Poiseuille flow between parallel walls and restrict the plot of both the normalized reduced density $\eta/\eta_0$ and the velocity $u_y$ on the right half $[0,\, L_c]$ of the computational domain.
In this figure, we can observe again the very good agreement between the FDLB and the PM results for $\eta_0=0.01$. For the larger value of $\eta_0$, the density profile has a similar behavior as in the case of dense gas at rest between two parallel walls in Sec. \ref{sec:gas_at_rest}, while the FDLB velocity profile is within a few percent from the PM one. The results are similar for both values of the confinement ratio used $R\in\{4,10\}$.

\subsubsection{\label{sec:large_a} Non-linear regime }

\begin{figure}
\includegraphics[width=0.49\linewidth]{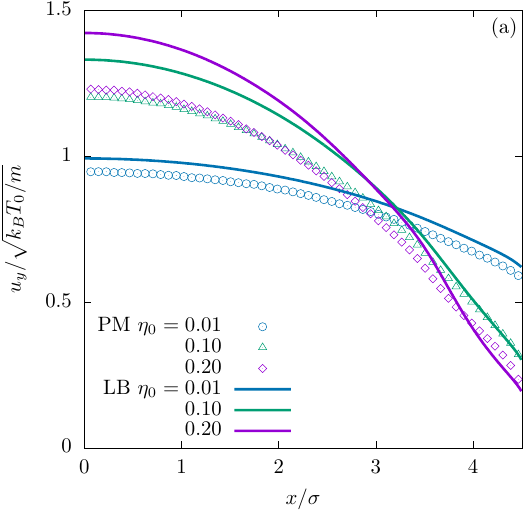}\hfill
\includegraphics[width=0.495\linewidth]{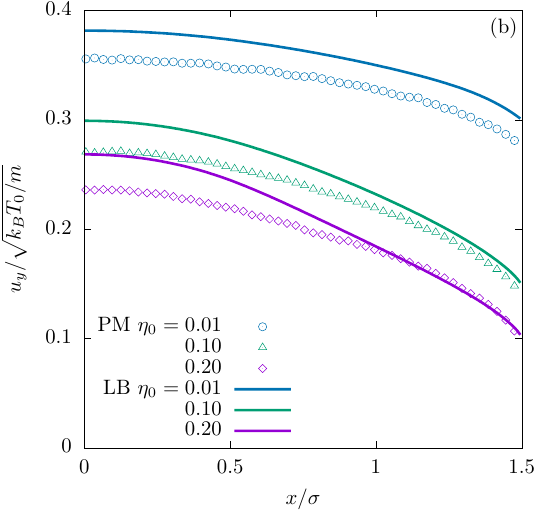}\\
\includegraphics[width=0.485\linewidth]{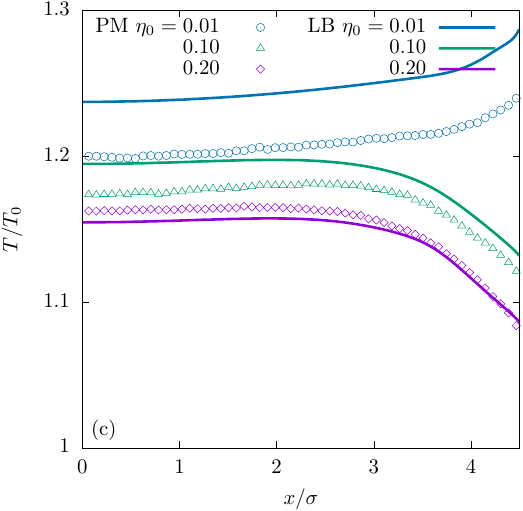}\hfill
\includegraphics[width=0.51\linewidth]{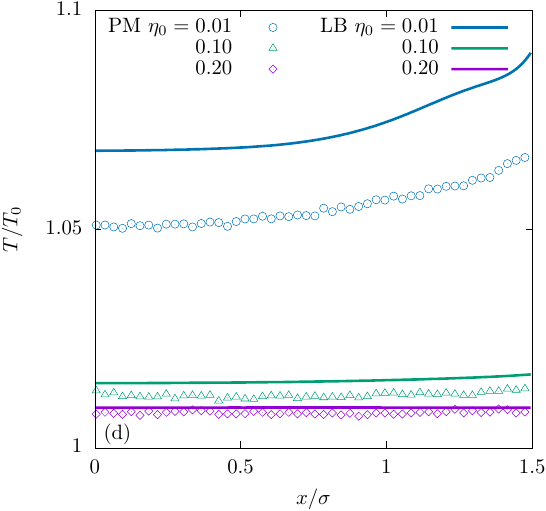}
\caption{\label{fig:poisseuille_vel_temp_a01} Poiseuille flow (non-linear regime): Velocity and temperature profiles at an external acceleration of $a_y=0.1$, three values of the mean reduced density $\eta_0=\{0.01,0.1,0.2\}$ and $R=10$ (left column) and $R=4$ (right column).}
\end{figure}

\begin{figure}
\includegraphics[width=0.49\linewidth]{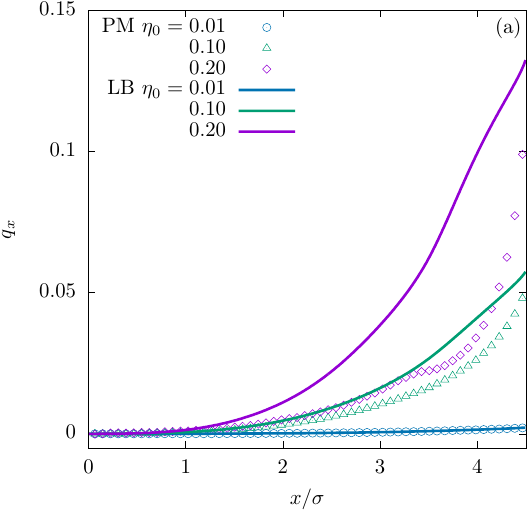}\hfill
\includegraphics[width=0.505\linewidth]{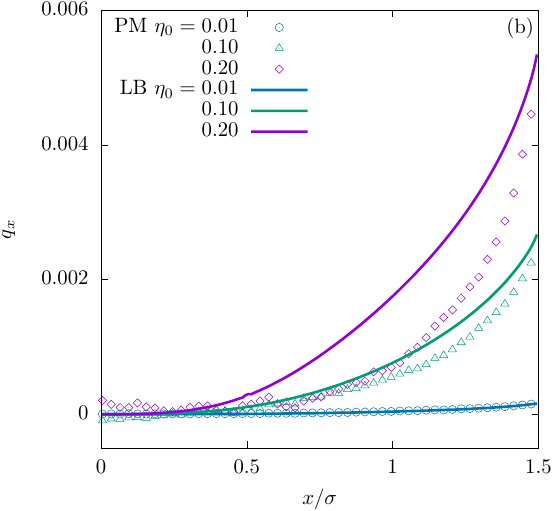}\\
\includegraphics[width=0.49\linewidth]{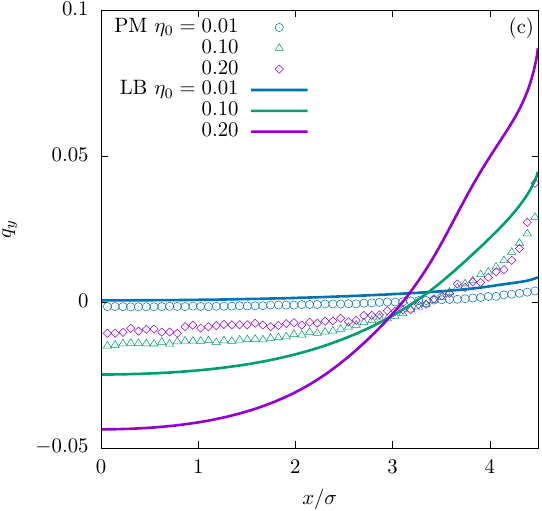}\hfill
\includegraphics[width=0.495\linewidth]{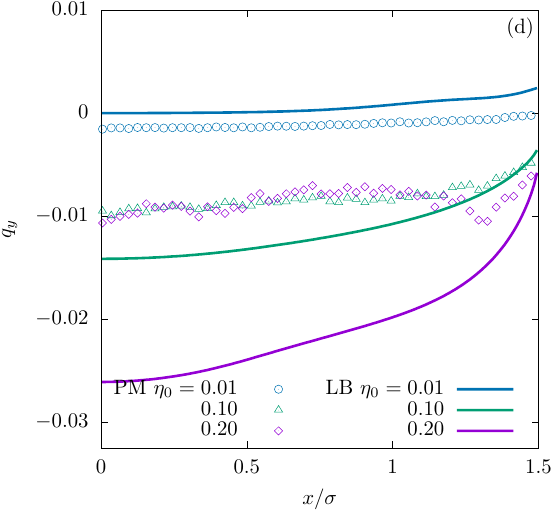}
\caption{\label{fig:poisseuille_qx_a01} Poiseuille flow (non-linear regime): Transversal and longitudinal heat flux, $q_x$ and $q_y$, at an external acceleration of $a_y=0.1$, three values of the mean reduced density $\eta_0=\{0.01,0.1,0.2\}$ and $R=10$ (left column) and $R=4$ (right column).}
\end{figure}

Moving to high acceleration of $a_y=0.1$, we also encounter transversal variations in both the temperature and the heat fluxes. As in Couette flow, in addition to the transverse component $q_x$ of the heat flux, the Poiseuille flow displays a non-zero longitudinal heat flux $q_y$, which arises as a distinct microfluidics phenomenon.

Fig. \ref{fig:poisseuille_vel_temp_a01} plots the velocity and the temperature profiles for the acceleration $a_y=0.1$. The density profile has insignificant variations with respect to the small acceleration case (see Fig. \ref{fig:poisseuille_dens_a0001}). Concerning the velocity profiles, the FDLB results are in good agreement with the PM curves, with larger deviations observed in the center of the channel as the reduced density $\eta_0$ is increased. Instead, a surprisingly good match between the FDLB and the PM profiles is obtained near the walls. Also, the temperature profiles are again in good agreement: the FDLB profiles exhibit a deviation less than $5\%$ with respect to the PM profiles throughout the channel.

In Fig. \ref{fig:poisseuille_qx_a01} we present the profiles of both the transversal and the longitudinal heat fluxes. An excellent agreement between FDLB and PM profiles is seen for the transversal heat flux at low values of the reduced density $\eta_0$. As seen also in the previous plots, at larger values of $\eta_0$ the deviations are huge, but this is an expected outcome due to the crude approximations used in the simplified Enskog collision operator. In the case of the longitudinal heat flux, the discrepancies are even more pronounced. Nevertheless, the model still provides a reasonably accurate description of the flow at low fluid densities.

\subsubsection{\label{sec:mfr} Mass flow rate}

The mass flow rate (MFR) is defined as:
\begin{equation}
 \dot{m}=\int_{-\frac{L_c}{2}}^{\frac{L_c}{2}}\frac{n(x)u_y(x)}{a} dx
\end{equation}
 and it was normalized by $\dot{m}_0=n_0 m a L_c^2/v_m$, where $v_m=\sqrt{k_BT_w/m}$.
The results are plotted in Fig. \ref{fig:poisseuille_mfr}(a), where we plot the normalized mass flow rate with respect to the Knudsen number for four values of the confinement ratio $R\in\{2,5,10,20\}$.
The tightness of wall confinement significantly affects the mass flow rate in the Enskog equation. When the channel width is large ($R>20$), the MFR at Knudsen numbers ($Kn$) larger than 1 mostly matches that of the Boltzmann equation. However, for $Kn < 1$, the MFR in the Enskog equation is smaller than predicted by the Boltzmann equation, and this difference becomes more pronounced as $Kn$ decreases. As the channel width decreases, the MFR becomes smaller, and the deviation from the Boltzmann equation extends to larger Knudsen numbers. For $R > 5$, the MFR does not monotonically decrease with $Kn$ in the slip flow regime ($Kn < 0.1$). Instead, there exists a specific value of $Kn$ at which the MFR locally reaches a maximum. When $R$ is less than or equal to 5, the Knudsen minimum disappears, and the MFR only increases with the Knudsen number. Following the study in Ref.~\cite{CLBG22}, the Knudsen minimum disappearance can be explained as follows. Under tight geometries, the combined contribution of viscosity and density weakens, while the slip term remains constant as the confinement ratio R decreases. As a result, the relative importance of the slip term increases in this context. Consequently, the disappearance of the Knudsen minimum under tight confinement can be attributed to the more pronounced significance of fluid slippage at the wall.

In Fig.~\ref{fig:poisseuille_mfr}(b), the normalized mass flow rate $\dot{m}/\dot{m}_0$ is evaluated for different quadrature orders $Q_x^{h}$ in the FDLB model with respect to the Knudsen number. Specifically, four values of the quadrature order $Q_x^h\in\{8,16,32,64\}$ were considered. This plot illustrates that, as the Knudsen number $Kn$ increases, one has to employ a larger quadrature order in order to approach the PM results. The results presented in Fig.~\ref{fig:poisseuille_mfr}(a) are obtained using a varying quadrature order over the interval of the Knudsen number. More specifically, a quadrature order of order $Q_x^h=8$ is sufficient for $Kn<1$ and then we switched to a quadrature of $Q_x^h=100$ to ensure the accuracy.

\begin{figure}
\includegraphics[width=0.5\linewidth]{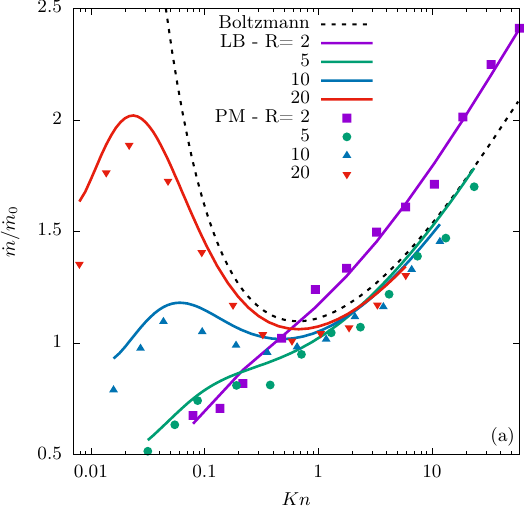}\hfill
\includegraphics[width=0.5\linewidth]{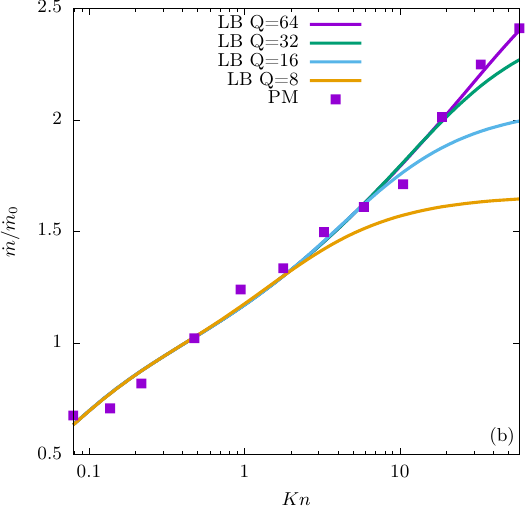}
\caption{\label{fig:poisseuille_mfr} Poiseuille flow (mass flow rate): (a) Normalised mass flow rate $\dot{m}/\dot{m}_0$ at an external acceleration of $a_y=0.001$ with respect to Knudsen number $Kn$, four values of the confinement ratio $R=\{2,5,10,20\}$. (b) Normalized mass flow rate $\dot{m}/\dot{m}_0$ with respect to the Knudsen number for four values of the quadrature order $Q_x\in\{8,16,32,64\}$. One can observe that as the Knudsen number increases one has to employ a higher order quadrature in order to recover the mean flow rate, as is the case for dilute gases.}
\end{figure}

\section{Conclusions}\label{sec:conclusions}\

In this work, a series of dense gas flows bounded by parallel plates were simulated in order to validate the proposed finite-difference Lattice Boltzmann model employing the simplified Enskog collision integral. In this model, the Enskog collision integral is approximated using a Taylor expansion and retaining the first-order gradients. We benchmarked the model in the following setups: Fourier flow, Couette flow, and Poiseuille flow.
The simulation parameters range from a low reduced density value ($\eta_0=0.01$) to a relatively high value ($\eta_0=0.2$) and include the confinement ratios of $R\in\{4,10\}$.

The FDLB results obtained for the Couette flow, Fourier flow, and Poiseuille flow were validated against the corresponding PM results. Reasonable agreement was observed throughout the parameter range. More specifically, our kinetic model adequately captures the effects of denseness, density inhomogeneity, as well as nonequilibrium phenomena within the range of flow parameters investigated.

In the case of the Fourier flow, we further examined the transversal heat flux with respect to the Knudsen number $Kn$ where we observed a significant deviation from the dilute gases results. Specifically, the heat flux is no longer monotonic and moreover exhibits a local maximum, whose value for $\eta_0=0.2$ is even larger than the ballistic limit. This is attributed to the potential contributions to the heat flux, as detailed in Appendix~\ref{appendix:heat_flux}.

The Couette flow results are well captured by the FDLB model, even if the deviations with respect to the PM results become larger as the fluid is more confined~(i.e. smaller $R$). Besides the density, velocity, and temperature profiles, we have also presented the transversal and longitudinal heat flux profiles. Since the simplified Enskog collision operator approximation discards the higher-order contributions to the collisional momentum and energy transfer, the FDLB results for the transversal and the longitudinal heat fluxes agree with the PM results only at very low reduced density ($\eta_0=0.01$).

As in the case of the Fourier and Couette flow, also the Poiseuille flow results are in good agreement with the PM results up to a mean reduced density of $\eta_0=0.1$. As shown before~\cite{WLRZ16,SGLBZ20,CLBG22}, the Knudsen minimum in the Poiseuille setup disappears for ultra-tight confinement ($R<5$). This effect is well captured by the FDLB model.

In conclusion, the presented model demonstrates its capability in handling moderately dense gases. Additionally, we examined the model's performance in dealing with flows characterized by sharp gradients in macroscopic quantities arising in the gas-surface interaction. This was achieved with a much lower computational cost compared to the particle method simulations.
Moving forward, our future plans include incorporating attractive forces between molecules to address multiphase flows.

\begin{acknowledgments}
This work was supported through a grant from the Ministry of Research, Innovation and Digitization, CNCS - UEFISCDI, project number PN-III-P1-1.1-PD-2021-0216, within PNCDI III. The authors thank dr.~V.E. Ambrus for useful discussions regarding the present study.

\end{acknowledgments}

\appendix
\section{Full-range Gauss Hermite quadratures: construction and comparison to half-range GH quadrature results}\label{appendix:fullrange}

Let's examine integrals of $f$ and $f^{\mathrm{eq}}$ along the entire axis of the 1D momentum space:
\begin{equation}
 \begin{pmatrix}
 M_s\\
 M_s^{(\rm{eq})}\rule{0mm}{7mm}
 \end{pmatrix} = \int_{-\infty}^\infty dp\,
 \begin{pmatrix}
 f \\
 f^{\rm{eq}} \rule{0mm}{7mm}
 \end{pmatrix}
 p^s. \qquad \label{eq:mom1d_def}
\end{equation}
where $f^{eq} = n g$ and
\begin{equation}%\label{eq:maxwelliana}
g \equiv g(x, p, t) = \frac{1}{\sqrt{2\pi m T}}
 \exp\left[-\frac{(p - m u)^2}{2T}\right]. \label{eq:feq-full}
\end{equation}
The function $g$ can be expanded with respect to the full-range Hermite polynomials
${H_\ell(p), \ell = 0, 1, \dots}$ as follows~\cite{AS16b,AS16a}:
\begin{equation}
 g = \frac{\omega(\bar{p})}{p_0} \sum_{\ell = 0}^\infty \frac{1}{\ell!}
 \mathcal{G}_\ell H_\ell(\bar{p}),\qquad
 \mathcal{G}_\ell = \sum_{s = 0}^{\lfloor \ell / 2\rfloor} \frac{\ell!}{2^s s! (\ell - 2s)!}
 \left(\frac{mT}{p_0^2} - 1\right)^s \left(\frac{mu}{p_0}\right)^{\ell - 2s},\label{eq:g_H}
\end{equation}
where $\bar{p}\equiv p / p_0$ represents the particle momentum with respect to an arbitrary momentum scale $p_0$, and $\lfloor \cdot \rfloor$ denotes the floor function.

The full-range Hermite polynomials~\cite{AS16a,H87,SYC06} satisfy the following orthogonality relation with respect to the weight function $\omega(p)$:
\begin{equation}
 \int_{-\infty}^\infty dp\, \omega(p) H_\ell(\bar{p}) H_{\ell'} (\bar{p}) = \ell! \,
 \delta_{\ell,\ell'}, \qquad
 \omega(p) = \frac{1}{\sqrt{2\pi}} e^{-\bar{p}^2/2}.
 \label{eq:H_ortho}
\end{equation}

The expansion coefficients $\mathcal{G}_\ell$ given in Eq.~\eqref{eq:g_H} are obtained as follows:
\begin{equation}
 \mathcal{G}_\ell = \int_{-\infty}^\infty dp\, g\, H_\ell(p).
\end{equation}

By substituting Eq.~\eqref{eq:g_H} into Eq.~\eqref{eq:mom1d_def}, we obtain:
\begin{equation}
 M^{( \mathrm{eq} )}_{s} = p_0^s \sum_{\ell = 0}^\infty \frac{1}{\ell!} \,\mathcal{G}_\ell
 \int_{-\infty}^{\infty} dp \,\omega(p)\, H_\ell(p) \, p^s.
 \label{eq:mom_aux}
\end{equation}
For finite values of $s$ and $\ell$, the Gauss-Hermite quadrature can be applied to compute the integral over $p$ across the entire momentum axis using the following approach:
\begin{equation}
 \int_{-\infty}^\infty dp\, \omega(p) P_s(p) \simeq \sum_{k = 1}^Q w_k^H
 P_s(p_k),
 \label{eq:GH_def}
\end{equation}
where $P_s(p)$ is a polynomial of order $s$ in $p$, and the $Q$ quadrature points $p_k$ ($k = 1, 2, \dots Q$) are the roots of the Hermite polynomial of order $Q$, i.e., $H_Q(p_k) = 0$. The quadrature weights $w_k^H$ are given by:
\begin{equation}
 w_k^H = \frac{Q!}{[H_{Q + 1}(p_k)]^2}. \label{eq:H_wk}
\end{equation}
The equality in Eq.~\eqref{eq:GH_def} is exact if $2Q > s$.
In an LB simulation, $Q$ is fixed at runtime. Thus, in order to ensure the
exact recovery
of $M^{( \mathrm{eq} )}_s$ \eqref{eq:mom_aux}, the sum over $\ell$ in Eq.~\eqref{eq:g_H} must be truncated at a finite value $\ell = N$. Setting $Q > N$ ensures the exact recovery of the first $N + 1$ moments (i.e. $s = 0, 1, \dots N$) of $f^{eq}$, since the terms of higher order in the expansion of $g$ are orthogonal to all polynomials $P_s(p)$ of orders $0 \le s \le N$, due to the orthogonality relation \eqref{eq:H_ortho}. This allows us to obtain $M^{\mathrm{eq}}_s$ as follows:
\begin{equation}
 M^{( \mathrm{eq} )}_s = \sum_{k = 1}^Q f^{eq}_k \bar{p}_k^s,\qquad
 f^{eq}_k = n g_k^H, \qquad
 g_k^H = \frac{w_k^H p_0}{\omega(\bar{p}_k)} g^{H,(N)}(\bar{p}_k),
\end{equation}
where $\bar{p}_k = p_0 p_k$ represents the discrete momenta, and the notation $g^{H,(N)}(p)$ indicates that the polynomial expansion \eqref{eq:g_H} of $g(p)$ is truncated at order $\ell = N$ using the full-range Hermite polynomials. For clarity, we provide the expression for $g^H_k$~\cite{AS16b,AS16a} below:
\begin{equation}
 g^{H,(N)}_k = w^H_k \sum_{\ell = 0}^N H_\ell(\bar{p}_k)
 \sum_{s = 0}^{\lfloor \ell/2 \rfloor} \frac{1}{2^s s! (\ell - 2s)!} \left(\frac{mT}{p_0^2} - 1\right)^s
 \left(\frac{mu}{p_{0}}\right)^{\ell -2s}.
 \label{eq:H_gk}
\end{equation}

In the case of the Poiseuille flow, the momentum derivative $\partial_{p_y} f$ can be written as:
\begin{equation}
 (\partial_{p_y} f)_k = \sum_{k,k'} \mathcal{K}_{k,k'} f_{k'},
\end{equation}
where the kernel $\mathcal{K}_{k,k'}$ has the following components
~\cite{ASFB17arxiv,BA19}:
\begin{equation}
 \mathcal{K}_{k,k'} = -\frac{w_k^H}{p_0} \sum_{\ell = 0}^{Q -2} \frac{1}{\ell!}
 H_{\ell + 1}(\bar{p}_k) H_{\ell}(\bar{p}_{k'}).\label{eq:kernel_H}
\end{equation}

In Fig.\ref{fig:full_vs_half} we plot the results obtained using the Full-range Hermite polynomials against the half-range results for $R=10$ and $\Delta T=0.5$, $u_w=1$ and $a_y=0.001$, for Fourier, Couette and Poiseuille flows, respectively. The reduced density was set to $\eta_0=0.01$.
We can easily see that a significantly larger velocity set needs to be employed in the case of Full-range polynomials in order to obtain similar results to those obtained using a half-range quadrature of $Q^h_x=8$, with 16 velocities. In most cases, a quadrature of $Q^H_x=200$ is needed in order to have relative errors of less than $1\%$.

\begin{figure}
\includegraphics[width=0.32\linewidth]{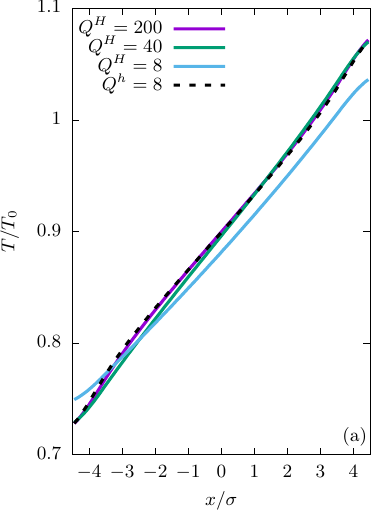}
\includegraphics[width=0.32\linewidth]{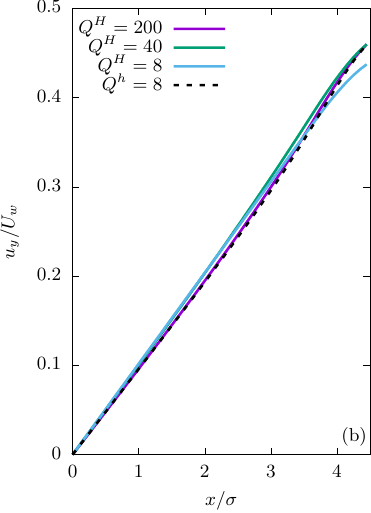}
\includegraphics[width=0.34\linewidth]{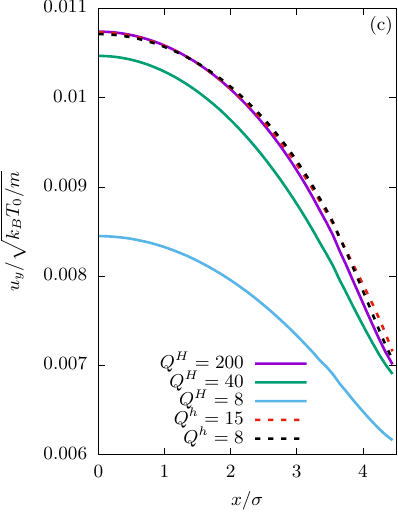}
\caption{\label{fig:full_vs_half} Comparison of half and fll-range Hermite polynomials: (a) Fourier flow (b) Couette flow and (c) Poiseuille flow at $R=10$ and $\Delta T=0.5$, $u_w=1$ and $a_y=0.001$, while the reduced density was set to $\eta_0=0.01$. These plots indicate a convergence of the results obtained using full-range quadrature towards those obtained using half-range quadrature. }
\end{figure}

\section{\label{appendix:heat_flux} Local maximum in heat flux for the Fourier flow }

Further looking at the heat flux in the context of the Fourier flow, one can evaluate the kinetic and potential contributions to the total heat flux. While the kinetic component is evaluated as usual, the potential part is given by~\cite{F99}:
\begin{multline}
 q_x^{pot}=-m\frac{\sigma^2}{4}\int d\bm{\xi}_1 d \bm{\xi}_2 d^2 \bm{k} \int_0^\sigma d \alpha ({\xi_1^*}^2-\xi_1^2) \bm{k} \chi\left({\bm{x}}+\alpha \bm{k}-\frac{\sigma}{2}{\bm{k}}\right) \\
 f({\bm{x}} +\alpha \bm{k}-\sigma {\bm{k}},\bm{\xi}_1)f({\bm{x}+\alpha\bm{k}},\bm{\xi}_2)({\bm{p_r}}\cdot{\bm{k}})
\end{multline}
where $\alpha$ is a dummy variable.

Using the particle method described below one may assess the individual contributions to the total heat flux.
In Fig.~\ref{fig:heat_flux_explicit} we plot the components of the kinetic $\widetilde{q_x^{kin}}=\int_{-\frac{L_c}{2}}^{\frac{L_c}{2}} q_x^{kin} dx$ and potential $\widetilde{q_x^{pot}}=\int_{-\frac{L_c}{2}}^{\frac{L_c}{2}} q_x^{pot}dx$ heat flux contributions integrated over the channel length with respect to the Knudsen number. The total heat flux $q_x=q_x^{kin}+q_x^{pot}$ is also plotted in this figure, being the same curve corresponding to $\eta_0=0.2$ as in Fig.~\ref{fig:termtrans_qx}(b). One can easily observe the monotonic increase in the kinetic component while the potential contribution has two local maxima in the interval $Kn\in\{1:10\}$. The total heat flux has a value larger than the free molecular limit due to the potential contribution to the heat flux.

\begin{figure}
 \includegraphics[width=0.5\linewidth]{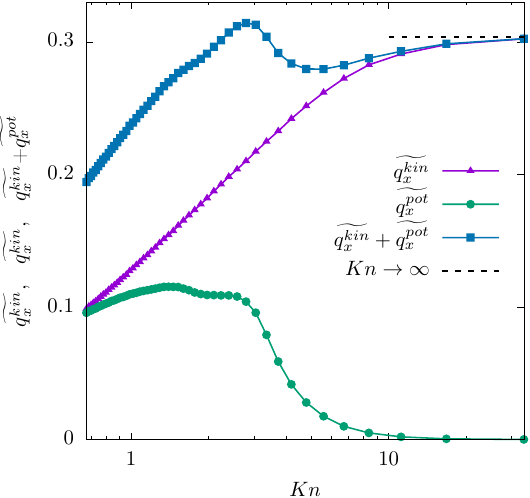}
 \caption{Heat flux: The components of the heat flux, kinetic $q_x^{kin}$ and potential $q_x^{pot}$, with respect to the Knudsen number. The kinetic component is monotonic with respect to the Knudsen number, while the potential part has two local maxima between $Kn=1$ and 10, leading to a total heat flux that exceeds the free molecular limit.}\label{fig:heat_flux_explicit}
\end{figure}

\section{\label{appendix:pm} Particle method for the Enskog equation}

The Enskog equation (Eq.\eqref{eq:enskog}) is numerically solved using a particle method that extends the original Direct Simulation Monte-Carlo (DSMC) method to handle the nonlocal nature of the Enskog collision integral~\cite{F97b}. For a comprehensive explanation of the numerical scheme and an analysis of its computational complexity, please refer to Ref.~\cite{FBG19}. Here, we provide a brief overview of the scheme.

The DSMC framework used to solve the Boltzmann equation is maintained, with modifications made to the collision algorithm to accommodate the nonlocal structure of the Enskog collision operator. The distribution function is represented by $N$ computational particles:
\begin{equation}
f(\bm{x},\bm{p},t)= \frac{1}{m} \sum_{i=1}^{N} \delta{\left(\bm{x}-\bm{x}_i(t)\right)} \delta(\bm{p}-\bm{p}_i(t)),
\end{equation}
where $\bm{x}_i$ and $\bm{p}_i$ are the positions and momenta of the $i$th particle at time $t$, respectively. The distribution function $f(\bm{x},\bm{p},t)$ is updated using a fractional-step method that splits the evolution operator into two sub-steps: free streaming and collision. In the first stage, the distribution function is advanced from $t$ to $t+\Delta t$ by neglecting particle collisions, i.e., by solving the equation:
\begin{equation}\label{eq:stage_I}
 \frac{\partial f}{\partial t} +\frac{\bm{p}}{m}\cdot\nabla_{\bm{x}}f + \bm{F}\cdot\bm{\nabla}_{\bm{p}} f =0,
\end{equation}
which leads to updating the positions of the computational particles as follows:
\begin{subequations}
	\begin{align}
	\bm{x}_i(t+\Delta t)&=\bm{x}_i(t)+\frac{\bm{p}_i}{m}\Delta t+\frac{F}{m}\frac{(\Delta t)^2}{2},\\
	\bm{v}_i(t+\Delta t)&=\bm{v}_i(t)+\frac{F}{m}\Delta t.
	\end{align}
\end{subequations}
resulting in the updated distribution function denoted $\tilde{f}(\bm{x},\bm{p},t+\Delta t)$.

In the second stage, the short-range hard-sphere interactions are evaluated, and the distribution function is updated according to:
\begin{equation}
 f(\bm{x},\bm{p},t+\Delta t)= \tilde{f}(\bm{x},\bm{p},t+\Delta t)+J_E[\tilde{f}]\Delta t.
\end{equation}
During this stage, the $N$ particle positions $\bm{x}_i$ remain unchanged, while their momenta $\bm{p}_i/m$ are modified based on stochastic rules that essentially involve the Monte Carlo evaluation of the collision integral given by Eq.~\eqref{eq:collision_integral} by selecting collision pairs accordingly.

The macroscopic quantities are obtained by averaging the microscopic states of the particles over time and performing phase averaging by running statistically independent simulations with identical macroscopic initial conditions but different random seeds.

\bibliography{bibliography.bib}

\end{document}